\documentclass[9pt,twocolumn]{extarticle}
\usepackage{preamble}

\title{\papertitle}
\author{%
  Koen van Greevenbroek\textsuperscript{1,*},
  Steven J. Davis\textsuperscript{1},
  Ken Caldeira\textsuperscript{1,2}
  \\[0.4em]
  \small\textsuperscript{1}Department of Earth System Science, Stanford University, Stanford, CA, USA\\
  \small\textsuperscript{2}Gates Ventures, Kirkland, WA, USA\\
  \small\textsuperscript{*}Corresponding author: \href{mailto:koenvg@stanford.edu}{koenvg@stanford.edu}
}
\date{}

\begin{document}

\begin{abstract}
The global food system contributes to tens of millions of years of human life lost every year through unhealthy diets\cite{afshin-sur-ea-2019}, and accounts for roughly a third of greenhouse gas (GHG) emissions\cite{crippa-solazzo-ea-2021}.
Researchers have thus often sought opportunities to simultaneously redress these large impacts on human health and climate; ``dietary shifts'' are frequently identified as key to both improving health\cite{rockstrom-thilsted-ea-2025,afshin-sur-ea-2019,bechthold-boeing-ea-2017} and reducing land-related GHG emissions\cite{stehfest-bouwman-ea-2009,clark-domingo-ea-2020,eisen-brown-2022}.
Here we develop a new, spatially-explicit, detailed model of the global food system, and use it to highlight enormous and low-cost opportunities to both improve dietary health and reduce food system emissions. However, we find that health and climate goals are surprisingly independent; for example, reducing red meat intake benefits climate, but may only modestly improve health outcomes.
Similarly, there is very large potential both to achieve healthier diets that do not lead to emissions reductions and to reduce emissions \emph{without} dietary change.
Our results highlight both the potential for, and the separability of, food-related policies that advance public health and climate goals.
\end{abstract}

\maketitle


The current global food system is damaging both human health and the climate system.
Chronic diseases due to malnutrition and dietary risks cause trillions of dollars in damage each year \cite{ruggeriladerchi-lotze-campen-ea-2024}, and the Global Burden of Disease study\cite{hay-ong-ea-2025} attributes 151 million years of life lost to dietary risks in 2020.
Further, food-system-related emissions\cite{crippa-solazzo-ea-2021} amounted to roughly a third of anthropogenic GHG emissions in 2015, including \qty{11.7}{\giga\tonne\COtwoeq} of emissions from land management and land-use change.

These enormous, twin burdens have led researchers and policymakers to routinely link food system impacts on climate and public health \cite{clark-domingo-ea-2020,rockstrom-thilsted-ea-2025,ruggeriladerchi-lotze-campen-ea-2024,bodirsky-beier-ea-2025}. In particular, studies have emphasized the climate and health benefits of reducing animal-source foods, and especially red meat \cite{tilman-clark-2014,bouvard-loomis-ea-2015,chan-lau-ea-2011,poore-nemecek-2018,hayek-harwatt-ea-2020,eisen-brown-2022}.
This has shaped a broader narrative of Planetary Health-style diets as a joint health-and-climate strategy \cite{springmann-godfray-ea-2016,willett-rockstrom-ea-2019,rockstrom-thilsted-ea-2025,humpenoder-popp-ea-2024,springmann-wiebe-ea-2018,stehfest-bouwman-ea-2009,bodirsky-beier-ea-2025,springmann-clark-ea-2018}.
Elsewhere, however, studies have indicated that food system emissions can be substantially reduced through supply-side measures alone --- raising crop and livestock productivity, reallocating production across regions, and sparing agricultural land --- \emph{without} dietary change\cite{havlik-valin-ea-2014,roe-streck-ea-2019,costa-wollenberg-ea-2022,beyer-hua-ea-2022}.
Here, we seek to reconcile these seemingly divergent findings by systematically assessing the opportunities to reduce emissions and/or improve health outcomes by changing what people eat and where and how food is produced.

\begin{figure*}[t]
  \centering
  \begin{minipage}[c]{0.66\textwidth}%
    \centering
    \includegraphics[width=\linewidth]{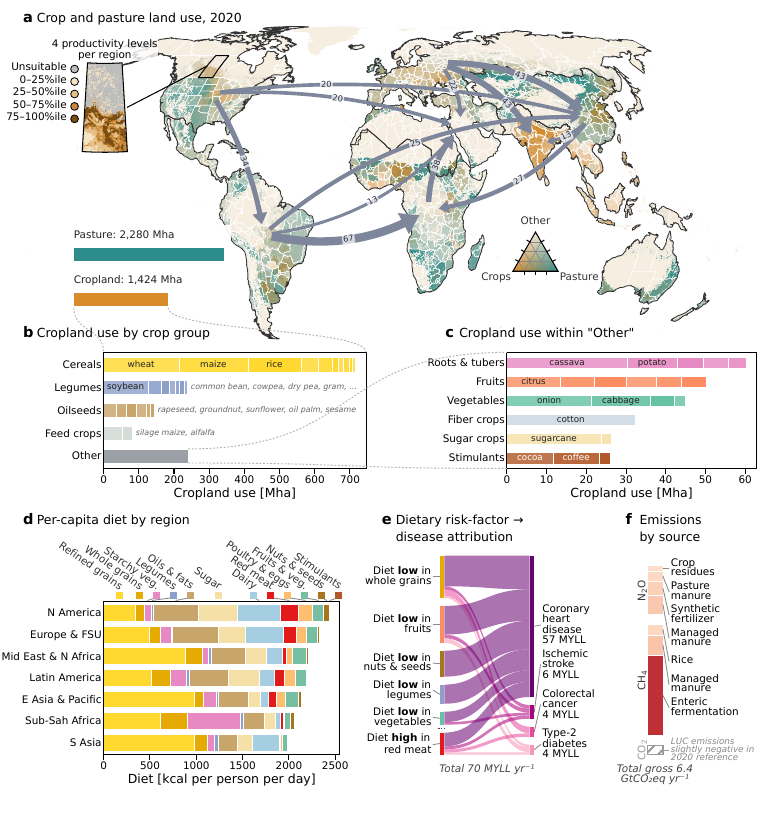}%
  \end{minipage}%
  \hfill
  \begin{minipage}[c]{0.32\textwidth}%
    \caption{%
      \textbf{Modeled reference global food system (2020).}
      \textbf{a},~Land-use intensity in the 2020 reference food system, shown as a ternary mix of cropland (orange), pasture (teal), and remaining unused or non-agricultural land (off-white) per cell; values are relative to total cell area.
      The inset map of Saskatchewan illustrates how each of the model's 750 regions is split into four productivity classes, and bars below the map show total 2020 cropland and pasture area in Mha.
      Subtle dark outlines delineate the seven UN-style super-regions (loosely UN M49), and arrows depict the largest inter-regional trade flows between them, annotated in Mha of land-equivalent at each arrow head (intra-region trade excluded; arrow width also scales with flow magnitude).
      \textbf{b},~2020 cropland use (Mha) by major crop group --- cereals, oilseeds, legumes and dedicated feed crops --- with everything else folded into an aggregated ``Other'' bar.
      Individual crops within each bar are stacked largest-first.
      \textbf{c},~Zoom into the ``Other'' bar of panel b, broken down into the remaining crop groups (roots and tubers, vegetables, fruits, sugar crops, stimulants, fiber crops).
      Bar colors in b and c match the corresponding food-group hue in panel d where applicable.
      \textbf{d},~Population-weighted per-capita food-group consumption (kcal per person per day) in the 2020 reference, by super-region, ordered top-to-bottom by total caloric intake.
      \textbf{e},~Sankey decomposition of the modeled dietary-risk health burden in the 2020 reference.
      Six dietary risk factors (left, gray) connect to four GBD diseases (right, colored) with ribbon width proportional to attributable years of life lost.
      Disease totals (in \unit{MYLL\per yr}) annotate the right column.
      \textbf{f},~Gross non-\ch{CO2} food system emissions in the 2020 reference (\ch{CH4} + \ch{N2O} in GWP\textsubscript{100}-equivalent \ch{CO2}), broken down by gas and by emission source.
      The panel omits the \ch{CO2} account (hatched stub): land-use-change clearing is negligible at the reference, but the cost-optimal solution leaves some existing agricultural land uncultivated and earns a spared-land regrowth credit of about \qty{\landSinkRef}{\giga\tonne\COtwoeq\per yr}, so net food-system emissions (\qty{\netEmRef}{\giga\tonne\COtwoeq\per yr}) fall below this gross total.%
    }%
    \label{fig:overview-map}%
  \end{minipage}%
\end{figure*}

Details of our model and analytic approach are provided in the Methods, and all code and data sources are extensively documented and openly available (see also \cref{fig:eda-topology}). In summary, we develop a new global food system model, GLADE (Global Land, Agriculture, Diet and Emissions), that uses mixed-integer linear programming to track and optimize production, trade, and consumption of 48 crops and 7 animal products among 750 regions globally (roughly equal-area aggregations of province-level administrative regions), and resolves agricultural lands within regions at 5 arc-minute (\textasciitilde\qty{9}{km}) resolution and grouped into 4 productivity classes (\cref{fig:overview-map}). We then use this model to systematically explore changes in the global food system in response to different valuations for human health and climate. In contrast to existing food system / land use models such as MAgPIE~\cite{lotze-campen-muller-ea-2008,dietrich-bodirsky-ea-2019} and GLOBIOM~\cite{havlik-schneider-ea-2011}, GLADE directly co-optimizes health and food system emissions (see \cref{tab:eda-model-comparison} for a structural comparison) by monetizing a statistical value of a year of life lost (YLL) and a social cost of carbon, and including both total health burden and emissions in the model objective function in monetary terms.

More specifically, we apply protective and harmful relative risk factors from the Global Burden of Disease study~\cite{zheng-afshin-ea-2022,hay-ong-ea-2025} to translate levels of consumption of different food groups into years of life lost due to coronary heart disease, ischemic stroke, type-2 diabetes and colorectal cancer (\cref{fig:overview-map}, e). Different risk factors for the same disease are combined multiplicatively using a piece-wise linear approximation; per-capita caloric intake is kept fixed at per-country 2020 (reference case) levels throughout.

We model land-management (agricultural) and land-use change (LUC) emissions of \ch{CH4}, \ch{N2O} and \ch{CO2} from food production, aggregated using IPCC AR6 GWP\textsubscript{100} factors\cite{ipcc-2021} (27 for \ch{CH4}, 273 for \ch{N2O}), hereafter referred to as ``food system emissions'', and exclude forestry, fisheries, post-farm-gate processing and energy-related emissions.
\ch{CH4} from ruminant enteric fermentation is computed at a model region level using IPCC Tier~2 methodology\cite{dong-mangino-ea-2006,ipcc-2019} from feed dry-matter intake and feed-category-specific methane yields; additional \ch{CH4} sources are manure management and wetland rice paddies.
\ch{N2O} sources comprise direct and indirect emissions from fertilizer application, livestock manure and crop-residue decomposition, all using default IPCC Tier~1 factors~\cite{ipcc-2019}.
Land-use change \ch{CO2} fluxes are resolved on a 9~km grid by calculating the forest component of observed above-ground biomass~\cite{santoro-cartus-ea-2021} and soil organic carbon~\cite{poggio-desousa-ea-2021}; carbon stock changes resulting from land clearing are amortized linearly over a 30-year horizon.
Spared, reforested cropland or pasture results in negative emissions equal to the 30-year mean uptake of regenerating young forests~\cite{cook-patton-leavitt-ea-2020}, restricted to climatically forest-capable biomes~\cite{hayek-harwatt-ea-2020}; to keep reforestation geographically plausible, we cap any single country at returning no more than half of its current agricultural land to forest (Methods).

A least-cost optimum on its own reproduces neither today's diets nor today's geography of production, reflecting a complex web of consumer preferences, subsidies, tariffs and institutional frictions.
We therefore anchor both sides of the system to the 2020 reference through two calibrated terms in the objective (Methods). On the demand side, \emph{consumer values} give each food a diminishing marginal benefit, calibrated by revealed preference so that its value at baseline intake matches the 2020 marginal cost of supplying that food.
On the supply side, a \emph{deviation penalty} charges the optimizer for moving cropland, pasture and livestock feed away from their observed 2020 distribution. Both are tuned so that, at zero value of life and cost of carbon, the cost-minimal solution closely reproduces the reference global food system of 2020.

To monetize health impacts in our study, we explore values of 50--\qty{50000}{USD} per year of life lost (YLL).
We arrive at this range by considering commonly-used statistical values of life; roughly 0.1--\qty{1}{million\,USD} in low- and lower-middle-income countries and 6--\qty{11}{million\,USD} in high-income countries\cite{lindhjem-navrud-ea-2011,viscusi-masterman-2017,robinson-hammitt-ea-2019}.
Converting to a value of a year of life by dividing by the remaining life expectancy (about \qty{45}{years}) of a person at the world's median age~\cite{unitednations-2024} yields a range of roughly 2000--\qty{250000}{USD\per YLL}. There is no need to go above \qty{50000}{USD/YLL} in our study, because nearly all gains are achieved at this level.

For emissions, we test social costs of carbon from 5 to \qty{500}{USD\per\tonne\COtwoeq}, spanning the long-standing US Interagency Working Group central value of \qty{51}{USD\per\tonne\COtwoeq}~\cite{interagencyworkinggrouponsocialcostofgreenhousegases-2021} 
and the \qty{185}{USD\per\tonne\COtwoeq} mean (5--95\,\% range 44--\qty{413}{USD\per\tonne\COtwoeq} 
of recent comprehensive damage-function work~\cite{rennert-errickson-ea-2022}.
Both the health and emissions monetization terms are applied uniformly and globally.
As a central counterfactual, we explore a \qty{\centralYllValue}{USD\per YLL} health valuation together with a \qty{\centralGhgPrice}{USD\per\tonne\COtwoeq} cost of carbon, both at the conservative, low end of the cited ranges. 

Finally, we perform a global sensitivity analysis to evaluate the effects of uncertainty in key input parameters, including dietary risk factors, conversion efficiencies in the food system (e.g., loss and waste, crop yields, feed conversion), emissions factors and warming potentials, and the value of lost life and social cost of carbon, using Sobol indices~\cite{sobol-2001} to quantify the relative importance of specific parameters (see Extended Data Table 2). 

\begin{figure*}[tb]
  \centering
  \includegraphics[width=\textwidth]{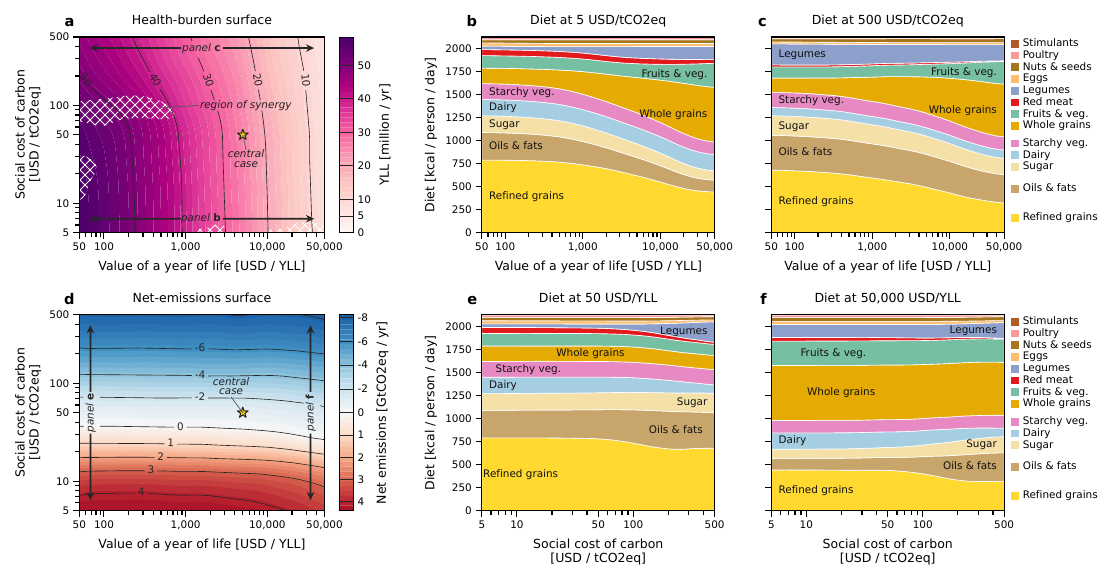}
  \caption{\textbf{Diet and health implications of pricing emissions and dietary health burden.}
    \textbf{a},~Total dietary health burden in YLL across a grid of combinations of the value of a year of life and the social cost of carbon, calculated as the median over a large sample of surrogate-model predictions trained on a global sensitivity analysis (Methods).
    The cross-hatched ``region of synergy'' marks the part of the (value of a year of life, social cost of carbon) grid where a \qty{10}{USD\per\tonne\COtwoeq} increase in the social cost of carbon lowers the median dietary health burden by at least \qty{0.5}{MYLL\per yr} --- our threshold for a meaningful marginal co-benefit; outside this region either the surface is approximately flat along the social-cost-of-carbon axis or the marginal effect of raising it further is too small to matter.
    \textbf{b}, \textbf{c},~Global population-weighted per-capita diet by food group as a function of the value of a year of life, at the low (\qty{\lowGhgRefPrice}{USD\per\tonne\COtwoeq}, \textbf{b}) and high (\qty{\highGhgRefPrice}{USD\per\tonne\COtwoeq}, \textbf{c}) edges of the social-cost-of-carbon grid.
    \textbf{d},~Net food system emissions (\ch{CO2} + \ch{CH4} + \ch{N2O} in GWP\textsubscript{100}-equivalent \ch{CO2}) on the same grid as \textbf{a}; the median is taken inside the Monte-Carlo ensemble so paired sequestration / emission draws keep their sign.
    Filled contours use a diverging colormap centered at zero.
    \textbf{e}, \textbf{f},~As \textbf{b}, \textbf{c}, but as a function of the social cost of carbon at the low (\qty{50}{USD\per YLL}, \textbf{e}) and high (\qty{50000}{USD\per YLL}, \textbf{f}) edges of the value-of-a-year-of-life grid.
    Solid labelled arrows along the edges of \textbf{a}, \textbf{d} mark the slice loci of the diet panels; the gold star marks the central healthy-and-sustainable operating point (YLL = \qty{\centralYllValue}{USD\per YLL}, carbon cost = \qty{\centralGhgPrice}{USD\per\tonne\COtwoeq}) cited throughout the paper.
    A decomposition of the dietary-risk health burden by attributable risk factor and GBD cause at four values of a year of life is shown in \cref{fig:eda-burden-decomposition}, fixed at the low social-cost-of-carbon slice for consistency with \textbf{b}.
    See also \cref{fig:eda-combined-sensitivity} for an explicit comparison of the value of a year of life and the social cost of carbon being introduced separately and jointly.}
  \label{fig:diet-health}
\end{figure*}

\subsection{The food system today}
In total, the dietary risks we model accounted for \qty{\yllRef}{MYLL\per yr} in 2020, with roughly four fifths attributable to coronary heart disease --- on the same order of magnitude as the global burden attributable to high alcohol use (\qty{39}{MYLL}) or tobacco smoking (\qty{126}{MYLL}) in 2020 (ref.~\refcite{hay-ong-ea-2025}).
Because we resolve only food-group risk factors --- holding other dietary risks such as high sodium intake fixed --- this is a conservative figure (GBD attributes \qty{151}{MYLL\per yr} to dietary risks across all causes in 2020; ref.~\refcite{hay-ong-ea-2025}), so if anything we understate the health gains achievable through dietary change (see Methods).
At a 2020 baseline, and not accounting for ongoing deforestation, we model food system emissions of \qty{\directEmRef}{\giga\tonne\COtwoeq\per yr} of direct \ch{CH4} and \ch{N2O} (\cref{fig:overview-map}f).

Producing this food costs about \qty{\prodCostRef}{bn\,USD\per yr} in direct production and trade outlays and occupies \qty{\croplandRef}{Mha} of cropland and \qty{\grasslandRef}{Mha} of pasture (\cref{fig:overview-map}).
The system is also deeply interconnected through trade: expressed in land-equivalent units, roughly \qty{\tradedLandPctAg}{\%} of all agricultural land is embodied in food traded across international borders (\cref{fig:overview-map}a).

These costs leave out the health and climate damages the present system imposes.
At our central valuations (\qty{\centralYllValue}{USD\per YLL} and \qty{\centralGhgPrice}{USD\per\tonne\COtwoeq}, both at the conservative low end of the cited ranges), today's unpriced health and climate externalities together amount to roughly \qty{19}{\%} of direct food production costs (\cref{fig:eda-today-externality-ratio}); see \cref{fig:eda-cost-surfaces} for the full decomposition across the (value of a year of life, social cost of carbon) plane and against the optimized system.

\subsection{Limited synergies between health and climate}

Although dietary health burden and emissions can be reduced jointly, the broader claim that the two goals are generally synergistic \cite{springmann-godfray-ea-2016,rockstrom-thilsted-ea-2025,springmann-wiebe-ea-2018} holds only in a narrow region of the policy space.
When a year of lost life is valued below roughly \qty{1000}{USD\per YLL}, raising the social cost of carbon does deliver a modest health co-benefit: animal-source foods and oils give way to legumes and, to a lesser extent, whole grains (\cref{fig:diet-health}a, e). Over this low-valuation range, lifting the carbon cost from \qty{\lowGhgRefPrice}{} to \qty{\highGhgRefPrice}{USD\per\tonne\COtwoeq} cuts median dietary-risk burden from about \qty{\yllSynergyLowGhg}{} to \qty{\yllSynergyHighGhg}{MYLL\per yr}, easing three modeled risk factors --- mostly through lower red meat intake, with the remainder from greater legume and whole-grain consumption (\cref{fig:eda-burden-attribution}).
At the highest social cost of carbon, more than half of the residual health burden is related to low fruit and vegetable consumption.

As the assumed value of a year of life increases, however, health and climate objectives decouple.
When a year of life is valued above \qty{\centralYllValue}{USD\per YLL} (i.e.\ at the lower end of the per-year values implied by commonly cited statistical values of life) regardless of the assumed cost of carbon, dietary health burden falls below \qty{\yllCeilingHighVal}{MYLL} (\cref{fig:diet-health}a).

The converse is also true --- and starker: the optimal emissions abatement as the assumed cost of carbon increases is largely insensitive to how a year of life is valued (\cref{fig:diet-health}d).
While social resistance to changes in agricultural practices might make it challenging to achieve these outcomes in the real world, net-zero food system emissions are cost-optimal at a social cost of carbon of only \netZeroGhgLo--\qty{\netZeroGhgHi}{USD\per\tonne\COtwoeq} across the full thousand-fold range of life valuation we explore (50--\qty{50000}{USD\per YLL}), and a social cost of carbon above \qty{\highGhgThreshold}{USD\per\tonne\COtwoeq} delivers net-negative emissions of around \qty{\highGhgEmCeilingAbs}{\giga\tonne\COtwoeq\per yr} or more at any value of life.

Climate and health outcomes are thus orthogonal above modest health valuations: any social cost of carbon lowers emissions and any valuation of lost life lowers health burden, but outside a region of synergy in which values of lost life are low, the two interact weakly. Within this region of synergy, emissions reduction and health improvement are driven mainly by reduced red meat intake, with increased legume consumption contributing the remainder.

The relative independence also means that health and climate outcomes can be improved in parallel without coming into conflict.
In a modest ``central'' case, valuing a year of life lost at only \qty{\centralYllValue}{USD\per YLL} and greenhouse gas emissions at \qty{\centralGhgPrice}{USD\per\tonne\COtwoeq}, a cost-optimal global food system still cuts the dietary health burden by \qty{\yllReducPct}{\%}, to about \qty{\yllCentral}{MYLL\per yr} while reducing net emissions by \qty{\emAvoided}{\giga\tonne\COtwoeq\per yr} from the zero-pricing reference (net \qty{\netEmRef}{\giga\tonne\COtwoeq\per yr}), to about \qty{\netEmCentral}{\giga\tonne\COtwoeq\per yr}.
Cropland and pasture contract by \qty{\croplandReducPct}{\%} and \qty{\grasslandReducPct}{\%} respectively, and direct production and trade costs fall from \qty{\prodCostRef}{bn\,USD\per yr} to \qty{\prodCostCentral}{bn\,USD\per yr}.
The spatial pattern of released cropland and pasture and the residual food-group burden at this central point are mapped in \cref{fig:eda-central-scenario-map}.

\subsection{Opportunities to reduce food system emissions}

\begin{figure*}[thb]
  \centering
  \includegraphics[width=\textwidth]{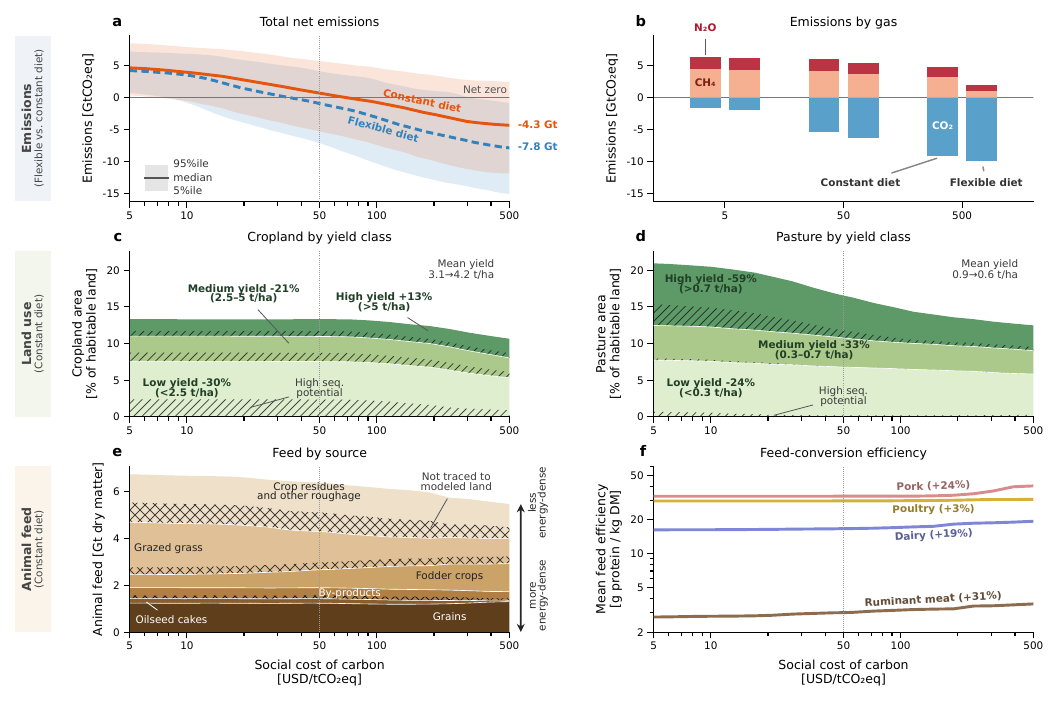}
  \caption{\textbf{Emissions abatement under flexible and constant diets.}
    All panels share the social cost of carbon on the x-axis (log scale); a faint vertical guide marks the central reference carbon cost (\qty{\centralGhgPrice}{USD\per\tonne\COtwoeq}), while the low and high reference carbon costs (\qty{\lowGhgRefPrice}{} and \qty{\highGhgRefPrice}{USD\per\tonne\COtwoeq}) coincide with the axis endpoints.
    \textbf{a},~Median net food system emissions as a function of the social cost of carbon, showing 5--95\% confidence intervals based on global sensitivity analysis (Methods) and comparing scenarios with flexible diets and constant human, but not animal, diets.
    \textbf{b},~Decomposition of net emissions by gas at the three reference carbon costs (\qty{\lowGhgRefPrice}{}, \qty{\centralGhgPrice}{} and \qty{\highGhgRefPrice}{USD\per\tonne\COtwoeq}), shown as paired stacked bars for the constant (left) and flexible (right) diet at each carbon cost; negative \ch{CO2} emissions correspond to carbon sequestration from reforestation.
    A further decomposition of the two non-\ch{CO2} gases by emission source, for both diets, is shown in \cref{fig:si-emissions-by-source}.
    \textbf{c}, \textbf{d},~Cropland \textbf{c} and pasture \textbf{d} area under a constant diet, expressed as a percentage of habitable global land area and split into low-, medium- and high-yield bands, showing increased utilization of high-yield cropland and a vast reduction in overall pasture area as the social cost of carbon rises.
    Hatching marks the sub-area on land with a high carbon sequestration potential, showing that primarily pasture with a high sequestration potential is released and reforested; the area-weighted mean yield trajectory is annotated in each panel.
    The full production-weighted yield distributions at three representative carbon costs are shown in \cref{fig:eda-yield-distributions}.
    \textbf{e},~Composition of global animal-feed intake by source category (\unit{\giga\tonne} dry matter) as a function of the social cost of carbon under a constant diet.
    The corresponding mosaic decomposition by animal class at three representative carbon costs is shown in \cref{fig:eda-feed-mosaic}.
    \textbf{f},~Production-weighted mean feed-conversion efficiency (g output protein per kg feed dry matter, the inverse of the feed-conversion ratio) per animal class as a function of the social cost of carbon.
    A higher social cost of carbon shifts production toward more feed-efficient systems, with the largest relative improvement for ruminant meat, the least feed-efficient class.}
  \label{fig:diet-abatement}
\end{figure*}

As previously noted, emissions reductions are relatively insensitive to health valuation, even though reduced red meat intake both improves dietary health and reduces emissions (\cref{fig:diet-health}).
In fact, the emissions reduction potential is largely insensitive to overall dietary composition, and substantial reductions can be achieved whether diets are allowed to shift (flexible) or not (constant).
Net-zero food system emissions are reached at a social cost of carbon of \qty{\netZeroGhgFlex}{USD\per\tonne\COtwoeq} if diets are flexible; holding diets fixed at the 2020 baseline roughly doubles the required carbon cost, to \qty{\netZeroGhgFixed}{USD\per\tonne\COtwoeq} --- still within the range of widely cited social-cost-of-carbon estimates.

More specifically, food system emissions can be substantially reduced without dietary changes by shifting crop and livestock production to higher-productivity regions, rebalancing the livestock feed mix and thus shrinking the land footprint of the global food system (\cref{fig:diet-abatement}).
When diets are constant, total crop production remains largely unchanged, but the area and production of pastures worldwide decrease substantially as ruminant livestock shifts away from bulky grass/forage feeds toward more energy-dense feeds and toward regions where management practices yield better feed conversion ratios; global animal-feed intake falls from \qty{\feedLowGhg}{Mt} to \qty{\feedHighGhg}{Mt} dry matter as a result (\cref{fig:diet-abatement}, e).
In turn, ruminant-meat production becomes markedly more feed-efficient: feed-conversion efficiency rises from about \qty{2.7}{\gram} to \qty{3.6}{\gram} output protein per kg feed dry matter (equivalent to a feed-conversion ratio falling from \fcrRumLowGhg{} to \fcrRumHighGhg{} kg feed dry matter per kg animal protein) as the social cost of carbon rises from \qty{\lowGhgRefPrice}{USD\per\tonne\COtwoeq} to \qty{\highGhgRefPrice}{USD\per\tonne\COtwoeq} (\cref{fig:diet-abatement}, f).

\subsection{Key uncertainties}

\begin{figure*}[thb]
  \centering
  \begin{minipage}[c]{0.72\textwidth}%
    \centering
    \includegraphics[width=\linewidth]{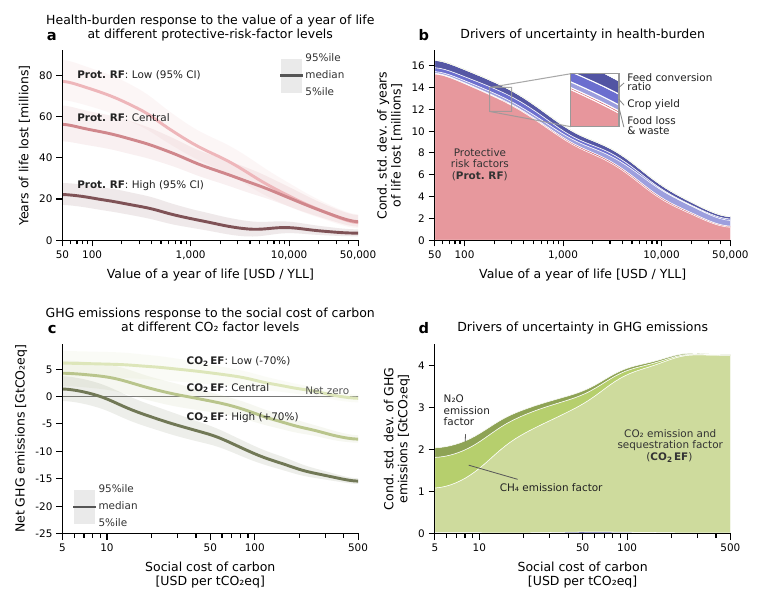}%
  \end{minipage}%
  \hfill
  \begin{minipage}[c]{0.26\textwidth}%
    \caption{\textbf{Key uncertainties.}
      Sensitivity of model results to uncertain, non-policy input parameters (\cref{tab:eda-uncertain-parameters}).
      In each panel the off-axis policy lever is fixed at its central value (\qty{\centralGhgPrice}{USD\per\tonne\COtwoeq}, \qty{\centralYllValue}{USD\per YLL}); the remaining parameters are sampled from their joint distribution.
      \textbf{a}, \textbf{c},~Median outcome trajectories with 5--95\,\% bands, conditioned on the dominant driver from the matching stack held at its low, central or high value --- years of life lost vs the value of a year of life for the protective dietary risk factors held at the low, central and high end of their GBD 95\,\% confidence interval \textbf{a}; net food system emissions vs the social cost of carbon for low, central and high CO\textsubscript{2} EF \textbf{c}.
      \textbf{b}, \textbf{d},~Per-parameter contribution to the conditional standard deviation of dietary-risk years of life lost, as a function of the value of a year of life \textbf{b}, and of total food system emissions, as a function of the social cost of carbon \textbf{d}; the bands stack to the total conditional standard deviation of the outcome.
      GHG uncertainty is dominated by the CO\textsubscript{2} emission and sequestration factor (\textbf{CO\textsubscript{2} EF}), largely the global sequestration potential of agricultural land; YLL uncertainty is driven primarily by the protective dietary relative-risk factors (\textbf{Prot. RF}) and the feed-conversion ratio.
      \Cref{fig:si-phase-sobol} extends this analysis from single-axis slices to the full (value of a year of life, social cost of carbon) plane.}%
    \label{fig:sobol-sensitivity}%
  \end{minipage}%
\end{figure*}

At a given value of a year of life or social cost of carbon, the standard deviation in total health burden as other parameters vary reflects epistemic uncertainty of optimal health outcomes. Similarly, the standard deviation of net food system emissions indicates epistemic uncertainty in climate outcomes.
Variation in the modeled global dietary health burden, in millions of years-of-life-lost, is most sensitive to the protective dietary relative-risk factors (\cref{fig:sobol-sensitivity}, a), with the baseline burden more than tripling between high and low estimates.
Animal-system feed conversion ratios are a minor secondary factor (\cref{fig:sobol-sensitivity}, b): more feed-efficient animal production frees feed-crop land for healthier human-food crops, lowering the cost of shifting toward a healthier diet and reducing the residual dietary-risk burden.
Food loss \& waste and crop yields exhibit the same dynamic, by changing the effective cost of food supply.

Similarly, variation in modeled food system emissions can be attributed almost entirely to uncertainty in land-use-change emission factors (\cref{fig:sobol-sensitivity}, c).
Across the parameter ranges we test, the standard deviation of total modeled food system emissions reaches \qty{\ghgConditionalStdMax}{\giga\tonne\COtwoeq} (\cref{fig:sobol-sensitivity}, d) --- comparable to the gross direct baseline emissions themselves (\qty{\directEmRef}{\giga\tonne\COtwoeq\per yr}).
This suggests that reducing uncertainty in the carbon sequestration potential of current agricultural land could substantially reduce the uncertainty in our modeled emissions.

\subsection{Health and climate as parallel policy objectives}

There is debate in public policy research about whether the distinct issues of health and climate should be linked narratively\cite{mayrhofer-gupta-2016} --- it is unclear, for example, whether highlighting potential health co-benefits of climate policy will make such policy more palatable\cite{maibach-nisbet-ea-2010,myers-nisbet-ea-2012} or whether bundling the two issues mainly alienates\cite{campbell-kay-2014}.
The actual existence and extent of any such co-benefits is a separate question; while promoted in the planetary health field \cite{horton-beaglehole-ea-2014,whitmee-haines-ea-2015}, progress toward both goals has been negatively correlated in the past\cite{pradhan-costa-ea-2017} due to links between health, wealth and consumption.
We argue that the two objectives are largely separable at a technical level.
Separate, parallel policies can act independently to improve dietary health (by targeting dietary change) and reduce food system emissions (by prioritizing reforestation).

We concur with previous studies\cite{tilman-clark-2014,springmann-godfray-ea-2016,willett-rockstrom-ea-2019,clark-domingo-ea-2020,searchinger-wirsenius-ea-2018} finding that reduced red meat intake is very advantageous from a climate perspective.
In practice, however, global per-capita meat consumption is projected to continue its increase \cite{oecd-fao-2025}, reduced meat intake is an unpopular climate solution\cite{maibach-nisbet-ea-2010} and alternative protein is yet to make serious inroads on the meat market.
In this context, our result that food system emissions can be reversed even without changing diets is encouraging.
Moreover, a separation of concerns can avoid alienation in polarized political environments.

While sociopolitical factors may make achieving these outcomes challenging, the potential we demonstrate for negative food system emissions (\qty{\emFixedHighGhg}{\giga\tonne\COtwoeq\per yr} at a social cost of carbon of \qty{\highGhgRefPrice}{USD\per\tonne\COtwoeq}) \emph{without} dietary change is a striking result --- previous studies\cite{havlik-valin-ea-2014,roe-streck-ea-2019,costa-wollenberg-ea-2022} with comparable scope have shown that interventions including yield improvements can reduce or halt deforestation and bring food system emissions close to net zero.
Other studies based on crop reallocation have demonstrated that total cropland could be reduced substantially\cite{hua-hu-ea-2025,beyer-hua-ea-2022,davis-rulli-ea-2017}, but do not quantify the yearly sequestration potential on current agricultural land under optimized scenarios.
In any case, endogenous optimization of the livestock sector and pasture land use in particular unlocks vast potential for emissions reductions that previous studies have failed to take advantage of. 

Reforestation of agricultural land, while having the potential to achieve net-zero food system emissions at a low cost, does come with significant local impacts.
We therefore cap the fraction of each country's current agricultural land that may be reforested at \qty{50}{\%}, while exploring sensitivity to this parameter in \cref{fig:eda-reforestation-surfaces}.
In our central scenario, which reaches net food system emissions of \qty{\netEmCentral}{\giga\tonne\COtwoeq\per yr} through \qty{\seqCentralAbs}{\giga\tonne\COtwoeq\per yr} of carbon sequestration on reforested land, this release is dominated by China and Brazil, which reforest \qty{\reforestedChnPctOwnAg}{\%} and \qty{\reforestedBraPctOwnAg}{\%} of their agricultural land respectively (almost all in the form of pasture); the remainder is spread across dozens of other countries, led by Russia and the USA.
By itself, this reforestation could threaten food sovereignty or significant agricultural export industries.
International carbon markets or equivalent mechanisms would be needed to ensure compensation --- in their current forms such markets struggle with over-crediting\cite{morita-matsumoto-2023} and spill-over effects\cite{meyfroidt-rudel-ea-2010,pendrill-persson-ea-2019}.

\subsection{Limitations}

We limit our investigation to the interactions between dietary health and food system emissions.
The food system has other impacts including eutrophication, biodiversity degradation and blue water use \cite{poore-nemecek-2018,mogollon-hadjikakou-ea-2026,tewierik-declerck-ea-2025}; these are outside the scope of the present study.
We also limit the scope of health burden under consideration to relative risk factors involving the consumption of major food groups.
These are the risk factors that relate directly to which kind of food is grown;
we do not in this study consider the health burden of oversupply or undersupply of calories (the high-body-mass-index and child-growth-failure risk factors, respectively) --- jointly accounting for roughly \qty{77}{MYLL} and \qty{86}{MYLL} globally in 2020 (ref.~\refcite{hay-ong-ea-2025}),
but rather keep calorie consumption constant at 2020 levels.
Neither omitted channel, however, has a strong link to food system emissions.
Emissions are dominated by land use, which scales with animal-source foods that supply less than a fifth of global calories (red meat only a few percent); the elasticity of food system emissions with respect to total calorie intake is therefore low, and reduced over-consumption would shift the climate ledger materially only insofar as it is routed through the red-meat channel we already capture.
Compositional shifts among plant foods --- the other route by which body-weight interventions act --- likewise move the climate ledger little, since land-use intensities vary far less among plant foods than between plant and animal sources\cite{poore-nemecek-2018}.
Child undernutrition is, in turn, largely a problem of distribution rather than aggregate supply.
Including these channels would expand the captured share of dietary-risk burden but leave the climate--health asymmetry that is our central finding largely intact.
Our study is static in time, considering counterfactual food systems around a 2020 reference year with fixed crop yields as well as populations.
The results therefore do not account for population growth, robustness against bad harvests or climate change --- these are interesting avenues for future research.

There is no behavioral model of demand: consumer values are calibrated to reproduce baseline consumption rather than fitted to observed price or income elasticities, and there are no cross-elasticities between food groups.
Sociopolitical resistance to large-scale changes is not explicitly modelled.
The counterfactual diets we report should therefore be read as cost-minimal scenarios bounding what is economically and biophysically possible, not as predictions of how consumers would respond to a given social cost of carbon or value of a year of life.
We model six of the food-group dietary risk factors identified by the Global Burden of Disease study~\cite{zheng-afshin-ea-2022,hay-ong-ea-2025} (whole grains, legumes, fruits, vegetables, nuts \& seeds and red meat) but neglect others --- notably high sodium and processed meat --- as well as micronutrient deficiency, obesity and child stunting / wasting.

\subsection{Conclusions}
Taken together, our findings reframe the relationship between dietary health and climate in the global food system.
Rather than searching for a single intervention with co-benefits for both objectives, our results suggest the two should be pursued as parallel opportunities.
Modest co-benefits do exist --- reduced red meat, together with increased legume consumption, simultaneously cuts emissions and dietary-risk burden --- but they operate only at very low health valuations.
At any health valuation in the policy-relevant range (above roughly \qty{2000}{USD\per YLL}), emissions policy delivers near its full mitigation regardless of health priorities, and a health-focused policy delivers near its full burden reduction regardless of the social cost of carbon.

Our findings suggest that health and climate objectives need to be pursued in parallel through complementary policy measures; we find limited potential for one policy to tackle both problems.
Nevertheless, the low valuations of life and emissions at which substantial health and climate gains become cost-optimal argue for the urgency of addressing dietary health and food-system decarbonization objectives.
In short, our study underlines the vast potential for parallel policies to cost-effectively advance food-related public health and climate goals.

\printbibliography[title={References}]

\clearpage
\begin{refsection}

\section{Methods}

\methodsfont

\subsection{Model overview}

GLADE is a global, cost-minimizing optimization model of the land-based food system.
It is written in Python, and the optimization structure is built using the PyPSA framework~\cite{brown-horsch-ea-2018}.
In its default configuration the model resolves 750 sub-national optimization regions obtained by clustering level-1 administrative units from the Global Administrative Areas database~\cite{gadm-2024}, \numCrops{} crops (11 cereals, 7 pulses, 5 roots and tubers, 4 vegetables, 6 fruits, 3 stimulants, 6 oil crops, 2 sugar crops, cotton, and 3 fodder/biomass crops), seven animal-product classes (bovine, pig, poultry and sheep/goat meat, cow and buffalo dairy, and eggs), and inter-regional trade.

The 750-region default is a deliberate compromise between sub-national heterogeneity in yields, climate and land-use change pressure on the one hand and solver tractability on the other; we verify in \cref{fig:si-region-resolution} that headline outcomes at the central healthy-and-sustainable operating point are insensitive to this choice, with the 750-region default agreeing with the highest-resolution (2000-region) run to within $\sim\!2$\,\% across net emissions, dietary-risk YLL, cropland and pasture.

Variables include land allocation to cropland and pasture (by productivity class and irrigation status), crop and animal production, food processing pathways, trade flows, and per-capita consumption; constraints enforce mass balance, resource availability (land, blue water, nitrogen), nutritional adequacy, and optional environmental ceilings.

The scope is the terrestrial food system: capture fisheries and aquaculture are not represented, and on the land-use side managed forestry is outside the model boundary --- forest enters only as a carbon-stock state that determines land-use-change emissions when cropland or pasture expands or contracts, with no endogenous timber, fuelwood or plantation activities.

The objective combines production, processing, trade, land-conversion, environmental and attributable-mortality costs (detailed later in this subsection), and is solved as a single-period problem with Gurobi~\cite{gurobioptimizationllc-2024}; the open-source HiGHS solver~\cite{huangfu-hall-2017} is also supported and is the default for users without a Gurobi license.

Because the piecewise-linear health-cost representation (see \nameref{sec:methods-health}) introduces binary segment selectors, the full problem is a mixed-integer linear program; with the health term disabled --- as in the constant-diet experiments of this paper --- it reduces to a pure linear program.

Individual solves complete in around 5\,minutes at the median when health costs are represented, with a tail to tens of minutes for the most policy-stressed cases, which makes the $\sim\!15{,}000$-solve uncertainty ensembles of this study feasible on standard high-performance computing infrastructure.

A complete listing of components, equations, unit conventions and configuration keys is provided in the online documentation at \url{https://sustainable-solutions-lab.github.io/GLADE/}.

The model shares its intellectual lineage with the global food-system optimization models MAgPIE~\cite{dietrich-bodirsky-ea-2019} and GLOBIOM~\cite{havlik-valin-ea-2014}, but differs from them in several ways that are material for the present study. First, the formulation is a single-period (mixed-integer) linear program, which trades some modeling fidelity (no price feedback outside the program, no period-by-period demand response) for tractability at high sampling rates. Second, both the code and the input-data build pipeline are fully open-source. Third, and most substantively, diet-attributable chronic-disease burden enters the optimization as an endogenous cost term rather than as a post-hoc diagnostic (see \nameref{sec:methods-health}). \Cref{tab:eda-model-comparison} summarizes the main structural differences; \cref{fig:eda-topology} shows the high-level commodity, emissions and policy-lever topology of the model.

\subsection{Supply chain representation}

The supply chain is organized as a directed flow network from primary resources to human consumption, with emissions accounted for at every transformation.

\paragraph{Land and land-use-change carbon.}
Cropland nodes are resolved per (region, productivity class, water-supply) triple and pasture nodes per (region, productivity class); cropland extents are initialized from the GAEZ~v5 downscaling of FAOSTAT 2019--2021 harvested area~\cite{fao/iiasa-2023}, pasture extents from ESA CCI Land Cover grassland fractions~\cite{copernicusclimatechangeservice-2019}, and the pasture stock is capped by the FAOSTAT permanent meadows-and-pastures area.
New agricultural land is drawn from a convertible-fraction pool distinguishing forest from non-forest origin; land-use-change \ch{CO2} emissions are computed at the grid cell from a sub-pixel decomposition of ESA Biomass CCI v6 above-ground biomass and SoilGrids~2.0 soil organic carbon~\cite{santoro-cartus-ea-2021,poggio-desousa-ea-2021}, and annualized over a 30-year amortization horizon.
Spared cropland and pasture earn sequestration credits using the Cook-Patton et al.\ global young-forest regrowth rates~\cite{cook-patton-leavitt-ea-2020} (a 30-year mean uptake, matching the land-clearing amortization horizon above), masked by the biome-eligibility classification of Hayek et al.~\cite{hayek-harwatt-ea-2020}.
To prevent implausibly concentrated land-use change, a per-country constraint caps the agricultural area (current cropland plus pasture) that any one country may return to forest at a configurable fraction of its 2020 total --- \qty{50}{\%} in our central configuration --- with a small additive buffer that grandfathers the structural-minimum sparing a few countries cannot avoid, so that tight caps remain feasible.
\Cref{fig:eda-reforestation-surfaces} shows how net food system emissions respond to this cap as it is varied from 5 to \qty{100}{\%}, for both flexible and constant diets.

\paragraph{Water and fertilizer.}
Blue-water availability is resolved at the river-basin scale from the Huang et al.\ monthly gridded irrigation withdrawals~\cite{huang-hejazi-ea-2018}, which represent current agricultural water use and are the variant used throughout this paper; a Water Footprint Network sustainable-availability variant~\cite{hoekstra-mekonnen-2011} can instead be selected by configuration.
Basin-level water budgets are distributed to regions using GAEZ-derived crop-specific irrigation requirements and growing-season dates, so that each region's share of a basin's water reflects both the timing and intensity of its irrigated cropping pattern.

Nitrogen fertilizer is modeled as a globally pooled resource at a marginal cost of \qty{500}{USD\per\tonne} of applied N; phosphorus and potassium enter implicitly through production costs.

\paragraph{Crop production.}
Crop yields for the \numCrops{} modeled crops are taken from the GAEZ~v5 \emph{actual yield} product (RES06-YLD), a 9~km downscaling of country-level FAOSTAT production and harvested area to gridded rainfed and irrigated yields, with input statistics taken as a three-year average over 2019--2021 (ref.~\refcite{fao/iiasa-2023}).
They are aggregated per (region, productivity class, irrigation status); productivity classes are yield-potential quantiles drawn from the GAEZ attainable-yield suitability layers and so preserve sub-regional heterogeneity without requiring gridded variables inside the optimization model.

Multi-cropping sequences from GAEZ agro-climatic zone data are treated as sequential within-year patterns; fertilizer and water requirements are summed across cycles.
Crop residues are partitioned between animal feed and soil incorporation; the latter carries an \ch{N2O} emission factor from incorporated residue nitrogen.

\paragraph{Non-food crops and biofuels.}
Two non-food uses share the cropland, water and nitrogen resources of the food system.
Cotton is included with its full processing chain --- lint to an externally-set fiber demand, anchored to FAOSTAT ``cotton lint, ginned'' production in the Crops and Livestock Products domain, with co-product cottonseed oil and cottonseed meal allocated through a single global ginning split~\cite{nationalcottoncouncilofamerica-2024} and fed to the model's vegetable-oil and protein-feed pools.

Biofuel and industrial-biomass demand is held at observed levels: liquid-biofuel and other industrial use are taken from the \emph{Other uses} element of the FAOSTAT Food Balance Sheets (FBS), drawn predominantly from maize grain and sugarcane (ethanol) and from vegetable oils (biodiesel) and routed via the corresponding food nodes, while silage-maize demand for anaerobic digestion in the major EU producers is curated from biogas-association statistics (see documentation for more detail).

In total, three dedicated biomass crops --- alfalfa, biomass sorghum and silage maize --- compete on equal footing with food and feed crops.
None of these non-food demands can be substituted with food within the model, so their demand quantities are fixed exogenously --- they still compete for land, water and nitrogen, but the policy levers act on the food system alone.

\paragraph{Livestock.}
Animal production is split into seven feed pools (ruminant roughage, forage, grain, and protein; monogastric low-quality, grain, and protein) to preserve the differences between grass-, concentrate- and by-product-based systems.
Pasture forage yields are taken from the ISIMIP2b LPJmL managed-grassland output~\cite{rolinski-muller-ea-2018} and scaled by a biomass-utilization fraction.
Country- and product-specific feed-conversion efficiencies are derived from GLEAM~3.0~\cite{herrero-havlik-ea-2013,fao-2024a} and the regional feed-energy ratios of Wirsenius~\cite{wirsenius-2000}.

Enteric \ch{CH4} and manure \ch{CH4} (VS~$\times$~B\textsubscript{0}~$\times$~MCF) are computed at IPCC Tier~2 (region- and product-specific factors) and manure \ch{N2O} (direct and indirect) at IPCC Tier~1 (default factors), following the 2019 Refinement~\cite{dong-mangino-ea-2006,ipcc-2019}.

\paragraph{Food processing, loss and waste, and trade.}
The model distinguishes between \emph{crops} --- primary agricultural commodities at the farm gate --- and \emph{foods}, the consumer-facing items that enter human diets and on which the nutrition and health modules operate.
We deliberately restrict the food layer to relatively lightly processed or unprocessed items (e.g.\ wheat flour rather than bread, refined sugar rather than confectionery), so the mapping from crops to foods is close to one-to-one for staples but branches where a single crop supplies several distinct food products: oil crops such as sunflower split into a whole-seed food and a refined vegetable-oil food, sugar crops into raw sugar and molasses, and cereals into flour, bran and germ streams.
In total the model resolves \numFoods{} food and processed-commodity nodes (some of which are co-products routed exclusively to feed or industry, e.g.\ oilseed meals, brans, distillers' grains) at a level of granularity that covers the global edible supply for the GBD dietary-risk factors, total caloric intake and global agricultural land use.
Processing is represented by multi-output conversion pathways with mass-balance factors (e.g.\ wheat~$\to$~flour~+~bran~+~germ), co-products are channeled to either feed or biomass export, and country- and food-group-specific loss and waste fractions derived from the SDG~12.3.1 indicator framework~\cite{unitednationsenvironmentprogramme-2024} are applied multiplicatively.

Inter-regional trade uses a hub network constructed by k-means on region centroids, reducing the number of potential links from quadratic to roughly linear in the number of regions while preserving the distance structure of global trade.
Every crop and food is assigned to one of four commodity classes --- bulk dry goods, bulky fresh produce, perishable high-value produce, and chilled meat --- with class-specific transport costs (between \qty{0.006}{USD\per\tonne\per\kilo\metre} for bulk dry grains and \qty{0.022}{USD\per\tonne\per\kilo\metre} for fragile fresh produce) charged on inter-hub trade links and scaling with great-circle distance.
The same class assignment carries a \emph{marketing cost} (30--\qty{200}{USD\per\tonne}) levied on the production or processing link, capturing the farm-to-wholesale handling, storage and first-handler margin that would otherwise be missing from a purely producer-price-based objective; the literature anchors for both cost parameters are documented online.

\paragraph{Nutrition and baseline diets.}
Nutritional composition of each of the modeled foods is drawn from USDA FoodData Central; we extract per-\qty{100}{g} energy, protein, carbohydrate and fat for every modeled food, but for the purposes of this paper only the caloric content enters the optimization actively, and total per-capita energy intake is held constant at the country level across every scenario, so that all diet shifts reported below are isocaloric reallocations across foods rather than changes in total food energy.

Every modeled food requires a 2020 baseline per-capita consumption value, which fixes the reference point against which the model is calibrated (see also \cref{fig:overview-map}) and serves as the anchor against which all scenario diets are compared.
Deriving this baseline is non-trivial: no single dataset reports intake at the granularity of the modeled foods on a globally consistent basis.
The construction has to balance three constraints simultaneously --- matching country-level total caloric intake, resolving food-group-level dietary survey data into individual modeled foods, and remaining compatible with the intake basis on which the GBD dietary-risk relative-risk curves are calibrated.

Baseline diets are built primarily from a per-country, per-food-group dietary-intake dataset for integrated-assessment use (GDD-IA) developed by M.\ Springmann \emph{et al.}\ (manuscript in preparation); it draws on the Global Dietary Database individual-level intakes~\cite{miller-singh-ea-2021} among other sources but is a separate harmonization product on a modeling-ready mass basis.
A particular strength of GDD-IA for our purposes is that it carries country-specific total caloric-intake estimates built up from anthropometric measurements (height and weight) combined with physiological equilibrium energy-balance modeling, rather than the under-reported self-report totals that bedevil survey-based intake studies; we treat these country-level kcal targets as the anchor onto which the per-food intakes are subsequently re-normalized (described below).
This is the same baseline diet dataset used in the EAT--Lancet~2.0 commission~\cite{rockstrom-thilsted-ea-2025}.

US group totals are taken from the NHANES~2017--2020 cycle to correct a known under-estimate of animal-source foods for the US.

For five of the six GBD dietary risk factors that enter the health module (fruits, vegetables, legumes, nuts and seeds, and whole grains), the GBD~2023 dietary-risk exposure estimate~\cite{hay-ong-ea-2025} replaces the GDD-IA group total wherever GBD reports a value, so the baseline aligns with the same intake basis the GBD relative-risk functions are calibrated against.
The sixth risk factor, red meat, is anchored on the supply side together with the other animal products instead (see below), so its baseline intake is set by FAOSTAT FBS supply rather than GBD exposure.

Within-group item shares are then allocated from FAOSTAT FBS supply, with country/global production-share blends used where several modeled foods share a single FBS item.

For the animal-product groups other than dairy (red meat, poultry, eggs), per-food intakes are instead anchored directly to FAOSTAT FBS supply, keeping baseline demand consistent with the slaughter-volume supply the model can physically deliver. The survey-based red-meat exposure we would otherwise carry exceeds this deliverable supply --- production net of feed, non-food and trade, after losses and waste --- at the global level, a physically infeasible baseline that FBS anchoring resolves. FBS supply also already encodes the trade, feed and stock-change accounting that the model must otherwise reconcile against, so anchoring to it keeps baseline demand and supply consistent at solve time.

Finally, because the preceding steps mix sources with different intake bases (survey-, GBD- and FBS-anchored, with cooked-to-raw conversions and group-to-food disaggregation along the way), per-food intakes can drift away from the country's caloric target.
We therefore close the construction with an anchor-aware kcal normalization: foods that have not been pinned by a GBD or FBS anchor are rescaled uniformly within each country until the total caloric intake of the baseline diet matches the GDD-IA kcal target derived above.
The result is a baseline diet that respects (i)~the GDD-IA caloric intake by country, (ii)~the GBD risk-factor intake basis for the five protective dietary risks, and (iii)~the FAOSTAT FBS supply-side accounting for animal products, including red meat.

\paragraph{Emissions accounting.}
\ch{CO2}, \ch{CH4} and \ch{N2O} are tracked on dedicated balance nodes and aggregated into a single greenhouse-gas account using IPCC~AR6 GWP\textsubscript{100} factors of 27 for \ch{CH4} and 273 for \ch{N2O}~\cite{ipcc-2021}; the \ch{CO2} account is restricted to land-use-change fluxes and the spared-land regrowth credit, and short-cycle biogenic \ch{CO2} from rice paddies and livestock respiration is excluded by convention (rice and livestock contribute only to the \ch{CH4} and \ch{N2O} accounts).

The model tracks emissions from the following sources, all of which feed into the three balance nodes above: (i)~enteric \ch{CH4} from ruminant fermentation, computed at IPCC Tier~2 from feed-intake and species/region-specific methane yields; (ii)~manure \ch{CH4} from manure management, also at IPCC Tier~2 via VS~$\times$~B\textsubscript{0}~$\times$~MCF; (iii)~direct and indirect manure \ch{N2O} from animal-housing and field application of organic N, at IPCC Tier~1 with the 2019 Refinement default factors~\cite{ipcc-2019}; (iv)~direct and indirect \ch{N2O} from synthetic-fertilizer application, with leaching, runoff and volatilization fractions parameterized as in IPCC Tier~1; (v)~\ch{N2O} from crop residues incorporated into the soil, scaled by residue-nitrogen content per crop; (vi)~rice-paddy \ch{CH4} from wetland-rice cultivation, with separate scaling for irrigated versus rainfed paddies; (vii)~land-use-change \ch{CO2} from conversion of forest and non-forest land to cropland or pasture, computed at the grid cell from the ESA Biomass CCI and SoilGrids decomposition described above and annualized over a 30-year amortization horizon; and (viii)~\ch{CO2} sequestration credits on land released back to natural cover, scaled by the Cook-Patton et al.\ regrowth rates and the Hayek et al.\ biome eligibility mask cited above.
Process emissions from downstream food processing, packaging, retail and household preparation are out of scope.

We retain GWP\textsubscript{100} rather than a flow--stock metric such as GWP*~\cite{allen-shine-ea-2018}: GWP* requires an emissions trajectory and is not naturally defined for the single-period counterfactuals modeled here, since applying it would require an assumed transition window between baseline and counterfactual that the model does not itself determine.
Were such a window imposed, GWP* would weight methane reductions more heavily than GWP\textsubscript{100}.

\paragraph{Objective function.}
The model objective minimizes the sum of (i)~\emph{crop production costs} derived from FAOSTAT Producer Prices, scaled by a share that excludes inputs already represented elsewhere in the objective (so feed, fertilizer and land costs are not double-counted); (ii)~\emph{livestock production and grazing costs} drawn from USDA ERS (US) and EU FADN accounts and value-allocated across joint products; (iii)~\emph{food-processing costs} on the crop-to-food conversion pathways and \emph{feed-conversion costs} routing crops, food co-products and residues into the seven feed pools; (iv)~\emph{land-conversion costs} differentiating forest ($\sim\!\qty{8000}{USD\per\hectare}$) from non-forest ($\sim\!\qty{2000}{USD\per\hectare}$) clearing, annualized at a 5\% discount rate over 30~years; (v)~\emph{trade and transport costs} differentiated by commodity category (bulk dry, bulky fresh, perishable high-value, chilled meat) and scaling with approximate distance through the trade network, together with the one-shot \emph{marketing cost} of 30--\qty{200}{USD\per\tonne} levied on the production/processing side of each commodity, as introduced above; (vi)~\emph{fertilizer costs} per tonne of applied nitrogen; (vii)~the \emph{deviation penalty} that charges the optimizer for moving cropland area, pasture area, and livestock feed composition away from their calibrated 2020 levels, representing the institutional and capital frictions that resist rapid reallocation (separately calibrated $\ell_1$ terms; see \nameref{sec:methods-calibration}); (viii)~a \emph{consumer-value} credit, entered with a negative sign, equal to a concave per-(country, food) utility of consumption whose marginal value at the 2020 baseline intake is calibrated to the baseline marginal supply cost (active in the flexible-diet configuration, where it anchors demand to observed diets; see \nameref{sec:methods-calibration}); and, as optional policy-lever terms, (ix)~a \emph{social-cost-of-carbon} term equal to the configured social cost of carbon (in \unit{USD\per\tonne\COtwoeq}) multiplied by total food system emissions, and (x)~an \emph{attributable-mortality} term equal to the configured value of a year of life multiplied by diet-attributable years of life lost (see \nameref{sec:methods-health}).

Further detail on cost derivation and unit conventions is given in the \textit{Costs} section of the online documentation.

\subsection{Health impact modeling}\label{sec:methods-health}

Endogenizing the diet-attributable chronic-disease burden lets us examine trade-offs between dietary health improvements and emissions reductions directly within the optimization.
The health cost is a monetized diet-attributable years-of-life-lost (YLL) term, built from the Institute for Health Metrics and Evaluation (IHME) Global Burden of Disease (GBD)~2023 release: dietary exposure--response (relative-risk) functions from the GBD~2023 Burden of Proof analysis (ref.~\refcite{zheng-afshin-ea-2022}), with dietary-risk exposure estimates and cause-specific baseline mortality rates from GBD~2023~\cite{hay-ong-ea-2025}.

\paragraph{Causes, risk factors, and population clusters.}
We model dietary impacts on the four non-communicable causes for which the GBD provides age-specific, cause-of-death dose--response curves linking food-group intake to disease risk: coronary heart disease (CHD), ischemic stroke (IS), type-2 diabetes mellitus (T2DM), and colorectal cancer (CRC).
Together, these four causes account for the large majority of the diet-attributable burden reported in GBD for the risk factors we model, and omit causes (e.g.\ respiratory and kidney diseases) for which GBD either does not report dose--response curves or does so only through channels (body mass index, blood pressure) that we do not resolve.
The six dietary risk factors that enter the module are fruits, vegetables, legumes, nuts and seeds, whole grains, and red meat.

Countries are grouped into 30 population clusters by geography, GDP-per-capita and population so the problem stays tractable.
For each cluster $c$ and cause $d \in \{\text{CHD, IS, T2DM, CRC}\}$, define
\begin{equation}
  \mathrm{YLL}_{c,d} \;=\; \sum_{a} M_{c,d,a}\, \bar{\ell}_{c,a},
  \label{eq:baseline-yll}
\end{equation}
the observed baseline cause-$d$ years of life lost in cluster $c$, where the sum runs over GBD age strata $a$, $M_{c,d,a}$ is the baseline cause-$d$ mortality count and $\bar{\ell}_{c,a}$ the corresponding remaining life expectancy.
This is the burden against which the model's diet-attributable share is computed.

\paragraph{Dose--response curves.}
Each risk factor $r$ (corresponding to a food group) acts on cause $d$ through a dose--response curve $\mathrm{RR}_{r,d}(x_r)$, fitted by the GBD, giving the cause-$d$ mortality hazard at intake $x_r$ relative to the minimum-risk reference intake of factor $r$ (specified precisely below, and distinct from the observed baseline diet).
The GBD~2023 Burden of Proof tool serves a single age-aggregated curve per risk--cause pair --- the risk at GBD's 60--64 reference age group, to which the estimated curve is assigned.
For the cardiovascular causes (CHD, IS), whose dietary effect attenuates with age, we restore an age-specific curve $\mathrm{RR}_{r,d,a}(x_r) = \exp\!\big(\beta_{d,a}\,\log \mathrm{RR}_{r,d}(x_r)\big)$ via a multiplicative log-RR attenuation factor $\beta_{d,a}$ whose age shape is taken from the GBD~2019 relative-risk appendix~\cite{murray-aravkin-ea-2020a} and normalized to the 60--64 reference age ($\beta_{d,a}\equiv 1$ for T2DM and CRC, which the GBD treats as age-invariant).
We then aggregate these age-specific curves into YLL-weighted effective per-cluster curves $\mathrm{RR}_{r,d,c}(x_r)$, so that the cluster-level curves reflect the demographic structure that determines how a given dietary shift translates into avoided deaths.

We take all six relative risk curves from the GBD~2023 Burden of Proof analysis (ref.~\refcite{zheng-afshin-ea-2022}) with one exception: for red meat we substitute a log-linear specification (constant relative-risk per \qty{100}{g\per\day} across the intake range), using per-outcome meta-analytic hazard ratios of \num{1.15} for CHD and \num{1.12} for IS from Bechthold~\emph{et al.}~\cite{bechthold-boeing-ea-2017}, \num{1.10} for T2DM from Li~\emph{et al.}~\cite{li-he-ea-2024}, and \num{1.17} for CRC from Chan~\emph{et al.}~\cite{chan-lau-ea-2011}.
The motivation is methodological rather than empirical: the GBD red-meat curve is steeply non-linear near zero intake, so the marginal benefit of removing the first gram of consumption is several times that of removing the hundredth.
In an optimization that can reallocate consumption across regions this disproportionately rewards eliminating already-low intakes in light-consuming populations over reducing the heavy intakes that account for most of the attributable burden; the log-linear specification keeps the per-\qty{100}{g} hazard ratio constant and removes this artefact.

\paragraph{From relative risk to attributable YLL.}
Multiple dietary risk factors are taken to act independently on each cause, so their cluster-level joint relative risk is the product of the factor-specific terms,
\begin{equation}
  \mathrm{RR}_{c,d}(\mathbf{x}) = \prod_{r} \mathrm{RR}_{r,d,c}(x_r).
  \label{eq:rr-product}
\end{equation}
The diet-attributable YLL at exposure $\mathbf{x}$ is then
\begin{equation}
  \mathrm{YLL}^{\mathrm{att}}_{c,d}(\mathbf{x}) = \mathrm{YLL}_{c,d}
    \cdot \frac{\mathrm{RR}_{c,d}(\mathbf{x}) - \mathrm{RR}_{c,d}(\mathbf{x}^{\mathrm{ref}})}{\mathrm{RR}_{c,d}(\mathbf{x}^{\mathrm{base}})},
  \label{eq:yll-att}
\end{equation}
where $\mathbf{x}^{\mathrm{ref}}$ is the theoretical-minimum-risk exposure level (TMREL), taken directly from the GBD~2023 risk-factor appendix, and $\mathbf{x}^{\mathrm{base}}$ is the observed baseline diet.
Note that $\mathrm{RR}_{c,d}(\mathbf{x}^{\mathrm{ref}})$ is by definition the minimum value of the risk curve over the considered intake range; usually $1$ for harmful risk factors and less than $1$ for protective factors.
The numerator measures the joint excess risk above TMREL at $\mathbf{x}$; normalizing by the \emph{constant} $\mathrm{RR}_{c,d}(\mathbf{x}^{\mathrm{base}})$ rather than by the canonical population-attributable-fraction denominator $\mathrm{RR}_{c,d}(\mathbf{x})$ keeps the expression linear in the joint relative risk and so embeddable in the optimization.
The two definitions coincide at $\mathbf{x} = \mathbf{x}^{\mathrm{base}}$ and track each other closely over the diet range of interest.

The contribution of cause $d$ in cluster $c$ to the objective is $V_{\mathrm{YLL}} \cdot \mathrm{YLL}^{\mathrm{att}}_{c,d}(\mathbf{x})$, with $V_{\mathrm{YLL}}$ the value placed on a year of life lost.
Note that, by definition, $\mathrm{YLL}^{\mathrm{att}}_{c,d}(\mathbf{x}^{\text{ref}}) = 0$, i.e.\ diet-attributable health burden as defined is exactly 0 when exposure to each risk factor (intake of each of the six food groups) is at the TMREL level.
Concretely, expressed in the model intake basis, these TMREL values are \qty{345}{g\per\day} for fruits, \qty{339}{g\per\day} for vegetables, \qty{185}{g\per\day} for whole grains, \qty{42}{g\per\day} for legumes, \qty{21.5}{g\per\day} for nuts and seeds, and \qty{0}{g\per\day} for red meat; each relative-risk curve is clipped at its TMREL so that intake beyond it yields no further benefit.
For the protective factors these reference intakes sit high in the observed exposure range, where the GBD dose--response curves flatten, so they reflect GBD's choice of minimum-risk anchor more than a careful physiological determination of an ``optimal'' intake; the fruit and vegetable references in particular are best read as conservative anchors rather than precise dietary targets.

\paragraph{Scope and conservatism of the modeled burden.}
The risk factors we resolve are a subset of those GBD evaluates: we include only food-group-level exposures and hold the remaining dietary risks (most importantly high sodium intake, but also fiber, trans-fatty acids, processed meat and sugar-sweetened beverages) fixed at their observed levels.
For 2020, GBD~2023 attributes \qty{151}{MYLL} to all dietary risks across all causes, and \qty{110}{MYLL} to the four causes we model (ref.~\refcite{hay-ong-ea-2025}); our reference baseline burden, given by \cref{eq:yll-att} evaluated at the observed diet and summed over clusters and causes, is \qty{\yllRef}{MYLL}.
The gap between \qty{110}{} and \qty{\yllRef}{MYLL} is the burden borne by the excluded dietary risks.
Because relative risks combine multiplicatively (\cref{eq:rr-product}), holding these factors fixed both drops their own contribution and reduces the leverage of the factors we do model.
Our baseline burden is reducible to zero only when every modeled intake reaches its TMREL, so it is a conservative estimate: the true health burden, and the benefit achievable through dietary change, are likely somewhat larger than what the model attributes to the six food groups alone.

\paragraph{Piecewise-linear embedding.}
Because \eqref{eq:rr-product} is a product of non-linear dose--response curves, $\mathrm{YLL}^{\mathrm{att}}_{c,d}$ cannot be embedded directly in a linear program.
We use the identity $\log(\prod_r \mathrm{RR}_{r,d,c}) = \sum_r \log \mathrm{RR}_{r,d,c}$ to convert multiplication to addition, and represent the two non-linear maps $x_r \mapsto \log \mathrm{RR}_{r,d,c}(x_r)$ and $z \mapsto \exp(z)$ by piecewise-linear interpolation on pre-computed breakpoint grids.

The inner stage $x_r \mapsto \log \mathrm{RR}_{r,d,c}(x_r)$, which is generally non-convex, uses a standard mixed-integer encoding in which exactly one segment of the piecewise-linear curve is active at a time; the outer stage $z \mapsto \exp(z)$, which is convex, is bounded below by its tangent lines (supporting hyperplanes at the breakpoints), an inequality that becomes tight at the optimum because the health cost rewards lower relative risk and so introduces no additional auxiliary variables.

We use 15 breakpoints on each stage; piecewise-linear error against direct non-linear evaluation is well below 1\% in baseline runs at this resolution, and degrades materially only below $\sim$10 breakpoints.

\subsection{Calibration, deviation penalty, and validation}
\label{sec:methods-calibration}

Gaps in input data sources, in particular pertaining to animal feed, mean that GLADE is unable to match crop and feed supply \& demand without some supply \& demand calibration.
We also employ a calibrated baseline deviation penalty in order to replicate current spatial production patterns.

\paragraph{Baseline calibration.}

We compute per-country downward yield and fodder-conversion corrections so that the observed animal feed balance (mainly derived from GLEAM data; see above) is reproducible under the FAOSTAT pasture-area cap; this addresses an apparent oversupply of forage feed in some countries given the assumed pasture area, yield and grazing intensity.
On the other hand, where ruminant forage demand exceeds what can plausibly be produced on current pasture at the assumed yields, the residual is supplied as an \emph{exogenous forage} stream at a low marginal cost.

Globally this exogenous / unaccounted-for supply amounts to about \qty{\exogForageGtdm}{\giga t~DM} per year and is spread across \exogForageCountries{} countries, with India, China, Pakistan, Brazil and Tanzania accounting for roughly \qty{\exogForageTopFivePct}{\percent} of the total.
Its presence reflects unresolved heterogeneity in pasture quality and in unreported forage sources (e.g.\ roadside grazing, leaves, crop-residue grazing, common lands), and it is held fixed across all scenarios so that it does not respond to policy levers.

Two further demand-side calibrations close residual supply--demand gaps that the raw input datasets leave open.
A \emph{food-waste calibration} adjusts the SDG~12.3.1 consumer loss-and-waste fractions~\cite{unitednationsenvironmentprogramme-2024} for the handful of food groups where the uniform global defaults materially mis-state the gap between FAOSTAT supply and GDD-IA intake (most strikingly vegetables, fruits and starchy vegetables), via per-food-group multipliers on the retained-share factor.
A complementary \emph{food-demand calibration} then applies per-food multiplicative adjustments to the GDD-IA baseline intake so that, after the waste correction, baseline supply and demand reconcile at the per-food level.
Both multipliers are extracted from the slack on a validation-mode solve (described below) in which mass-balance is closed by penalized slack generators on each food node, and both are clipped to bounded ranges so that they surface, rather than silently absorb, any large residual data inconsistencies.

The \emph{production-cost calibration} extracts additive per-(country, crop) and per-(country, product) cost adjustments from the shadow prices of a constrained solve in which the production allocation is held within $\pm 1$\% of the observed baseline.
This is to adjust our globally per-crop constant production costs for spatial heterogeneity and local subsidies.
Per-tonne cost adjustments are capped at a per-crop upper bound before the solve, and positive cost-calibration corrections are bounded above by the calibrated baseline so the calibration cannot drive unit production costs implausibly high.

\paragraph{Consumer values.}
With per-capita consumption free to vary within nutritional and food-group bounds, a cost-minimizing model does not reproduce observed diets: it substitutes the cheapest feasible foods for what people actually eat, subject only to the total caloric intake constraint.
We anchor demand to the observed 2020 diet through calibrated \emph{consumer values}, following a standard revealed-preference argument.
In a first baseline solve at zero social cost of carbon and zero value of a year of life, per-capita consumption of every food is fixed to its 2020 level (the GDD-IA-anchored baseline above); the dual variable of each per-(country, food) consumption equality then measures the marginal change in system cost per unit of relaxed consumption --- the marginal cost of supplying that food at its baseline quantity.
Provided observed intake sits near the equilibrium of supply cost and demand value, this dual is a proxy for the consumer's marginal willingness to pay at the baseline quantity.
The supply costs underlying it embed the subsidies, tariffs and frictions discussed below, so the recovered willingness to pay is the one revealed against today's distorted prices rather than a distortion-free structural preference; this suffices to anchor the baseline diet and bound the counterfactual demand response, but the calibrated values are conditional on the prevailing policy environment, which we hold fixed across scenarios.
We extract these duals --- flooring the occasional negative value, which signals a supply-side artefact (e.g.\ a forced co-product) rather than a preference --- and from each build a concave, piecewise-linear utility curve per country and food (four steps to either side of the baseline quantity, spanning zero to twice baseline intake, with the marginal value declining geometrically as consumption increases and matching the calibrated dual at the baseline anchor).
Entered into the objective as a benefit, this term makes the observed diet cost-optimal without pricing years of life lost or emissions and lets consumption depart from it in later scenarios only where health or emissions savings outweigh the forgone consumer value.

\paragraph{Deviation penalty.}
Without any anchoring to current production patterns, a model cost-minimization will deliver country-level baseline diets at a substantially lower aggregate cost than today, but from a sharply different production geography than the one we observe.
Today's distribution of crops and livestock is, in this sense, far from globally cost-optimal at current yields and feed-conversion ratios.
The gap reflects the dense web of subsidies and tariffs that distort producer incentives, together with cultural, political and practical frictions that resist rapid reallocation, none of which our model resolves explicitly.

Rather than attempting to reconstruct the underlying institutional and behavioral landscape country by country, we approximate its aggregate dampening effect with three global \emph{deviation penalty} parameters: $\lambda_c$ and $\lambda_p$ (both in bnUSD per Mha), $\ell_1$ penalties on per-country deviations of cropland and pasture area, respectively, from the calibrated 2020 allocation, and $\lambda_a$ (bnUSD per Mt DM), an $\ell_1$ penalty on per-country, per-feed-pool deviations of livestock feed composition from baseline.
Cropland and pasture carry separate coefficients because they react very differently to the same penalty: at current yields the optimizer finds large cost savings in reallocating cropland but only modest ones in reallocating pasture, so holding the two to a common deviation target requires a substantially stiffer penalty on cropland than on pasture.
All three enter the objective additively, so the optimizer remains free to respond to policy levers but pays a continuously rising price for reallocations far from today's system --- a role analogous to the gross-land-use-change and technological-change adjustment costs that anchor each period in the recursive-dynamic MAgPIE framework~\cite{dietrich-bodirsky-ea-2019}.

These coefficients are abstract modeling choices without an empirical basis.
We adopt a simple criterion for setting $\lambda_c, \lambda_p, \lambda_a$: they should be just strong enough that, at zero social cost of carbon and zero value of a year of life, the cost-minimal optimum deviates from baseline 2020 allocations by $5$\% for each of cropland area, pasture area and feed composition.
To find values satisfying this criterion we iteratively adjust the three coefficients with a quasi-Newton root-finder; the calibrated reference values are $\lambda_c = \qty{\devLambdaC}{bnUSD\per Mha}$, $\lambda_p = \qty{\devLambdaP}{bnUSD\per Mha}$ and $\lambda_a = \qty{\devLambdaA}{bnUSD\per Mt~DM}$, at which the zero-policy optimum realizes a \qty{\devCroplandPct}{\percent} cropland-area, \qty{\devPasturePct}{\percent} pasture-area and \qty{\devFeedPct}{\percent} feed-composition deviation from baseline.
We explore the sensitivity of our results to the parameter choice by rerunning the main analysis at values of $\lambda_c, \lambda_p, \lambda_a$ set half an order of magnitude above and below the reference values; see \cref{fig:si-stability-contour-surfaces,fig:si-combined-sensitivity-low,fig:si-combined-sensitivity-high}).

\paragraph{Validation.}
A separate \emph{validation} configuration fixes both per-capita consumption (to the GDD-IA-anchored baseline described above), per-country animal production (to the GLEAM 3.0 baseline) and observed crop harvested area (GAEZ, downscaled from FAOSTAT 2019--2021).
In this configuration, slack variables on food, feed and land nodes can close residual gaps between demand and supply at a high penalty cost.
In the default validation run land and feed slack are negligible (\qty{\valLandSlackMha}{Mha} on land, $\sim$\qty{e-3}{Mt~DM} on feed against demands of order \qty{1500}{Mha} and \qty{6800}{Mt~DM} respectively), and residual slack is concentrated on food-balance nodes at about \qty{\valFoodSlackPct}{\percent} of global food demand (\qty{\valFoodSlackMt}{Mt} in absolute value out of \qty{\valFoodDemandMt}{Mt}).
The largest residuals sit in dairy (\qty{\valDairySlackMt}{Mt}, $\sim$\qty{\valDairySlackOfDemandPct}{\percent} of dairy demand), vegetables (\qty{\valVegSlackMt}{Mt}) and fruits (\qty{\valFruitSlackMt}{Mt}); these are the food groups for which baseline FAOSTAT supply and GDD-IA intake are hardest to reconcile, and they are the same groups picked up by the food-waste calibration above.

\subsection{Experimental design}

Results in this paper derive from three complementary sets of model optimizations, each exploring a different dimension of the policy space.

\paragraph{The (social cost of carbon, YLL) pricing grid.}
Our principal exploration is a two-dimensional sweep across combinations of the social cost of carbon and the value of a year of life, run in the flexible-diet configuration of the model.
This grid frames the response-surface analysis in \cref{fig:diet-health} and underpins the headline claims about where the two policy objectives reinforce one another and where they decouple.
It is populated by the uncertainty ensembles described further below, with a surrogate emulator interpolating between the sampled points.

Within this grid we single out two deterministic single-point solves that anchor the headline numbers cited in the prose: a \emph{reference} run at zero social cost of carbon and zero value of a year of life, and a \emph{central} run at \qty{\centralGhgPrice}{USD\per\tonne\COtwoeq} and \qty{\centralYllValue}{USD\per YLL}.
The reference scenario is shown in the overview figure (\cref{fig:overview-map}).

\paragraph{Flexible- vs.\ constant-diet pairing along the social-cost-of-carbon axis.}
A second exploration contrasts \emph{flexible-diet} solves, in which per-capita consumption is free within nutritional and food-group bounds, against \emph{constant-diet} solves, in which consumption is pinned to the country-specific baseline.
We trace this contrast along the social-cost-of-carbon axis (\cref{fig:diet-abatement}) to isolate supply-side abatement channels, re-optimizing \emph{where} and \emph{how} food is produced, from demand-side channels, re-optimizing \emph{what} is eaten.
The years of life lost pricing lever is inactive in the constant-diet case (consumption is held constant, so the dietary-risk burden is too) and held at zero there.
A continuous deterministic sweep of constant-diet solves across the social cost of carbon provides the spatially resolved land-use and feed-mix trajectories of \cref{fig:diet-abatement} (rows 2 and 3); the abatement curves in the same figure are surrogate-median traces from the ensembles below.
The three representative solves at carbon costs of \qty{\lowGhgRefPrice}{}, \qty{\centralGhgPrice}{} and \qty{\highGhgRefPrice}{USD\per\tonne\COtwoeq} (low, central, high) drawn from this sweep underlie the production-weighted yield distributions (\cref{fig:eda-yield-distributions}) and the animal-feed mosaic (\cref{fig:eda-feed-mosaic}).

\paragraph{Uncertainty ensembles.}
For uncertainty quantification and variance decomposition we draw two scrambled Sobol ensembles over the uncertain parameters listed in \cref{tab:eda-uncertain-parameters}, with the policy levers held fixed on a grid at each draw so that the ensembles can be conditioned on policy at analysis time.

The \emph{flexible-diet} ensemble consists of $2^{14} = 16\,384$ samples over eight uncertain parameters; it is also replicated with $2^{11} = 2\,048$ samples at low and high deviation-penalty regimes as described under \nameref{sec:methods-calibration} (20\,480 solves in total).
This ensemble underlies the response-surface and sensitivity figures \cref{fig:diet-health,fig:diet-abatement,fig:sobol-sensitivity}, the burden decomposition and attribution panels (\cref{fig:eda-burden-decomposition,fig:eda-burden-attribution}), the deviation-penalty contour surfaces in \cref{fig:si-stability-contour-surfaces}, and the loose- and tight-penalty combined-sensitivity replicates in \cref{fig:si-combined-sensitivity-low,fig:si-combined-sensitivity-high}.

The \emph{constant-diet} ensemble of $2^{13} = 8\,192$ scenarios over the six non-health parameters, with the social cost of carbon as the only conditioning lever, underlies the constant-diet variant in \cref{fig:diet-abatement}.
In both ensembles the policy-lever parameters enter the objective linearly: the social cost of carbon multiplies the aggregated \ch{CO2}-equivalent emissions term, and the value of a year of life multiplies the attributable-mortality term.

\subsection{Uncertainty quantification and sensitivity analysis}

\paragraph{Parameters, distributions, and sources.}
We quantify uncertainty over six physical parameters --- crop-yield factor, \ch{CH4} emission-factor multiplier, \ch{N2O} emission-factor multiplier, land-use-change \ch{CO2} multiplier, food-loss-and-waste multiplier, feed-conversion-ratio multiplier --- and (in the flexible-diet ensemble only) two quantile parameters that interpolate within the GBD confidence bounds on the protective and harmful dietary relative-risk curves.
Ranges and distribution shapes are chosen from the most authoritative available uncertainty sources, and are listed together with their literature anchors in \cref{tab:eda-uncertain-parameters}.

\paragraph{Sampling and surrogate.}
Scenarios are drawn from a quasi-random (Sobol) sequence~\cite{saltelli-ratto-ea-2008} and mapped to the physical parameter ranges via their inverse cumulative distribution functions through \texttt{chaospy}~\cite{feinberg-langtangen-2015}.
On the resulting solved ensemble we fit a fast statistical emulator (a \emph{surrogate model}) that predicts the optimization outputs as a function of the full model input vector --- the uncertain parameters of \cref{tab:eda-uncertain-parameters} \emph{and} the policy levers (social cost of carbon and, in the flexible-diet ensemble, the value of a year of life), which all enter as surrogate features so that the emulator can be evaluated continuously across the joint uncertainty--policy space rather than only at the sampled policy grid: a multi-output XGBoost gradient-boosted-tree regressor~\cite{chen-guestrin-2016} (5\,000 boosting rounds, maximum tree depth~4, learning rate 0.02, early stopping on a 15\% Sobol-tail holdout) jointly fit on six aggregate scalar outputs (total cost, net \ch{CO2}, \ch{CH4}, \ch{N2O}, land use, and dietary-risk YLL) together with vector outputs for per-food-group consumption, per-feed-category mass, and dietary-risk YLL disaggregated by GBD cause and by risk factor.
Out-of-sample $R^2$ on the holdout exceeds 0.99 for the six aggregate scalar outputs in the flexible-diet ensemble (0.96 for the five non-YLL aggregates in the constant-diet ensemble, where YLL is structurally invariant), 0.97 for the YLL and feed-category disaggregations, and a median 0.97 for per-food consumption (tail to $\sim$0.8 for low-volume staples).
Three alternative surrogate families were tested --- polynomial chaos expansion, random forests, and multivariate adaptive regression splines --- but XGBoost offered the best accuracy--speed balance in development testing and is used throughout the paper.

\paragraph{Variance decomposition.}
First- and total-order Sobol sensitivity indices are computed on the fitted surrogate by Saltelli's pick-freeze estimator~\cite{saltelli-ratto-ea-2008} (a Monte-Carlo scheme that re-samples one parameter at a time to isolate its contribution to output variance) from a base sample of $N = 2^{12} = 4\,096$ Sobol points: the two independent $N \times D$ sample matrices and the $D$ one-column-swapped matrices are shared across all $D$ parameters, so each output's full set of first- and total-order indices is obtained from $N(D+2)$ surrogate evaluations.
Conditional indices along the policy-lever axes (\cref{fig:sobol-sensitivity}) are obtained by fixing the policy levers on a dense grid and re-sampling the remaining uncertain parameters.
Uncertainty bands in \cref{fig:diet-abatement} are the 5--95\% Monte-Carlo quantiles of the surrogate output with the two policy levers held fixed and all other uncertain parameters resampled; cross-regime dispersion from the three deviation-penalty replicates is reported separately as a robustness check.

\subsection*{Data availability}

All input datasets used by GLADE are public or freely registrable, with the exception of the GDD-IA baseline-diet product (in preparation; see below): harvested area, production, food-balance sheets and producer prices from FAOSTAT; gridded yields from GAEZ~v5~\cite{fao/iiasa-2023}; soil organic carbon from SoilGrids~2.0~\cite{poggio-desousa-ea-2021}; above-ground biomass from ESA Biomass CCI~v6~\cite{santoro-cartus-ea-2021}; land cover from ESA CCI~\cite{copernicusclimatechangeservice-2019}; administrative units from GADM~\cite{gadm-2024}; livestock parameters from GLEAM~3.0~\cite{herrero-havlik-ea-2013,fao-2024a}; dietary-risk exposure--response curves from the GBD~2023 Burden of Proof analysis~\cite{zheng-afshin-ea-2022}, with dietary-risk exposure estimates and cause-specific mortality from GBD~2023~\cite{hay-ong-ea-2025}; baseline diets from the GDD-IA dataset (Springmann \emph{et al.}, in preparation), building on the Global Dietary Database~\cite{miller-singh-ea-2021}.

The data underlying all figures and reported numbers, including per-figure CSV exports of all plotted arrays, are openly available in the paper's reproducibility repository (\url{https://github.com/koen-vg/reducing-health-climate-impacts-manuscript}) and archived on Zenodo (\url{https://doi.org/10.5281/zenodo.20618172}).
The GLADE model outputs required to regenerate those data --- solved networks, surrogate bundles, calibration artefacts, and analysis tables --- are deposited on Zenodo (\url{https://doi.org/10.5281/zenodo.20617942}).

\subsection*{Code availability}

The GLADE global food-systems optimization model is openly available at \url{https://github.com/Sustainable-Solutions-Lab/GLADE} under the GNU General Public License v3.0 (or later), with a pinned \texttt{pixi} environment and a \texttt{Snakemake}-based workflow~\cite{molder-jablonski-ea-2021} that reproduces all results in this paper from the inputs cited above; the specific release used for this study is archived on Zenodo (\url{https://doi.org/10.5281/zenodo.20618170}).

The code that builds the figures, tables, and prose-companion numbers from the model outputs --- together with the exact solve targets required to reproduce each figure --- is available in the reproducibility repository above (\url{https://github.com/koen-vg/reducing-health-climate-impacts-manuscript}) and likewise archived on Zenodo (\url{https://doi.org/10.5281/zenodo.20618172}).

Full documentation is available at \url{https://sustainable-solutions-lab.github.io/GLADE/}.


\printbibliography[title={Methods references}]


\section*{Acknowledgements}
The authors acknowledge use of large language models for drafting portions of the Methods section and for copy-editing the manuscript; all scientific content, results, and conclusions are the authors' own.

\section*{Funding}
This work was supported by an unrestricted gift from Gates Ventures LLC to Stanford University.

\section*{Author contributions}
K.v.G., S.J.D.\ and K.C.\ designed the study.
K.v.G.\ developed the GLADE model, performed the experiments and analyzed the results.
K.v.G.\ wrote the manuscript with input from S.J.D.\ and K.C.
All authors discussed the results and reviewed the manuscript.

\section*{Competing interests}
K.C.\ is employed by Gates Ventures LLC, which funded this work. The other authors declare no competing interests.

\section*{Additional information}
\noindent\textbf{Supplementary Information} is available for this paper.\\[0.2em]
\noindent\textbf{Correspondence and requests for materials} should be addressed to Koen van Greevenbroek (\href{mailto:koenvg@stanford.edu}{koenvg@stanford.edu}).

\clearpage
\begin{edtable}[!t]
  \section*{Extended Data}
  \vspace{-0.5ex}
  \centering
  \footnotesize
  \setlength{\tabcolsep}{4pt}
  \caption{\textbf{Structural comparison of GLADE with comparable global food-system models.}
    Rows report coarse characteristics rather than benchmark performance; for any given application, each model has strengths the others lack.}
  \label{tab:eda-model-comparison}
  \begin{tabular}{@{}llll@{}}
    \toprule
    Feature            & GLADE                      & MAgPIE                     & GLOBIOM                     \\
    \midrule
    Formulation        & (MI)LP                                 & Non-linear, recursive      & Non-linear, recursive       \\
    Spatial units      & $\sim$750 sub-national                 & $\sim$200 reg.\ $\times$ grid & 50 reg.\ $\times$ SimU    \\
    Time scope         & Single year                            & 1995--2100 recursive       & 2000--2100 recursive        \\
    Livestock detail   & Feed-pool (7 cat.)                     & Feed-basket                & Production-system grid      \\
    Diet response      & Endog.\ mix, calories fixed; surplus in obj. & Income regression; exog.\ targets opt. & Price-elastic (surplus)     \\
    Health cost        & Endogenous (GBD)                       & Post-hoc                   & Post-hoc                    \\
    Open-source        & Yes (code $+$ build)                   & Yes (code)                 & Partial                     \\
    \bottomrule
  \end{tabular}
\end{edtable}

\begin{edfigure}[!tbp]
  \centering
  \includegraphics{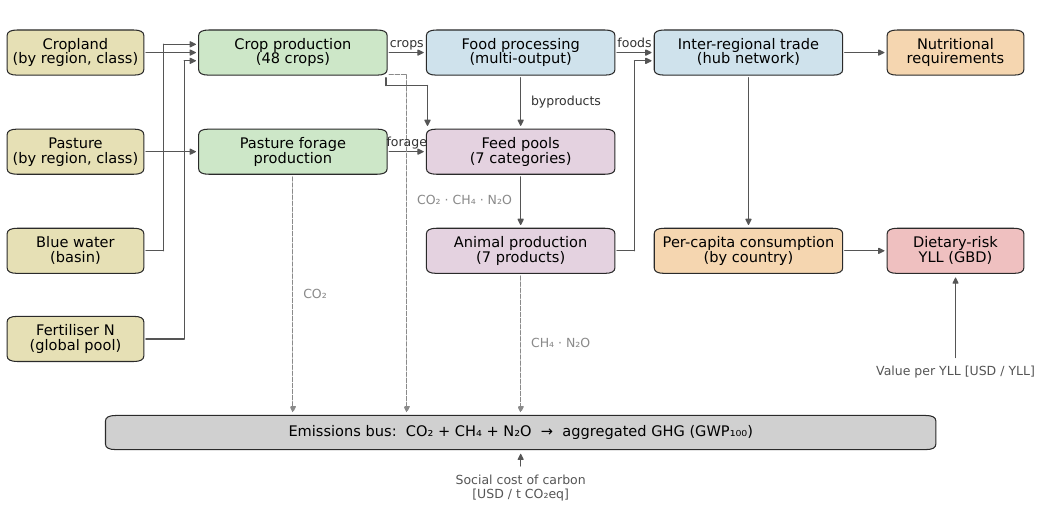}
  \caption{\textbf{High-level topology of GLADE.}
    Commodities flow from primary resources (cropland, pasture, blue water, fertilizer nitrogen) through crop and pasture forage production, food processing pathways (with processing byproducts routed to feed pools), feed pools and animal production, inter-regional trade, per-capita consumption, and nutritional outcomes.
    Emissions from land-use change, rice paddies, manure, fertilization, residue incorporation and enteric fermentation feed dedicated CO\textsubscript{2}, CH\textsubscript{4} and N\textsubscript{2}O buses that aggregate into a single greenhouse-gas account using GWP\textsubscript{100} factors (shown at the bottom).
    The two exogenous policy levers used in this paper --- the social cost of carbon and the value placed on a year of life lost --- enter the objective as scalar multipliers on the aggregated emissions and the diet-attributable-mortality terms respectively; their entry points are marked with open arrows.
    Each solid arrow corresponds to a class of multi-bus PyPSA links; the complete component inventory is given in the online documentation.}
  \label{fig:eda-topology}
\end{edfigure}

\begin{edfigure}[!tbp]
  \centering
  \includegraphics{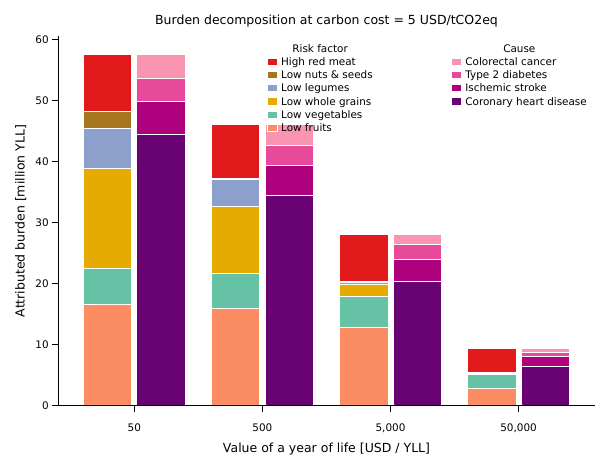}
  \caption{\textbf{Decomposition of dietary-risk health burden by attributable risk factor and GBD cause.}
    Companion to \cref{fig:diet-health}b.
    At a low social-cost-of-carbon slice (\qty{\lowGhgRefPrice}{USD\per\tonne\COtwoeq}, the lower edge of the displayed range, matching panel b of Fig.~2), the surrogate-median dietary-risk health burden is shown at four values of a year of life spanning the surrogate's range.
    Left bar of each pair: split by attributable risk factor, calculated by allocating each cause's burden across food groups in proportion to their excess log-relative-risk above the theoretical minimum-risk exposure level (TMREL), to account for the multiplicative nature of relative risk factors.
    Right bar of each pair: split by GBD cause (coronary heart disease, ischemic stroke, type-2 diabetes, colorectal cancer).
    Low intake of fruits and whole grains dominates the burden at low values of a year of life; low whole grains is largely eliminated by \qty{\centralYllValue}{USD\per YLL}, but low fruits persists as a major component of the (smaller) residual burden even at that valuation. High red meat intake is the most persistent risk factor and dominates the much smaller residual burden at the highest valuations.}
  \label{fig:eda-burden-decomposition}
\end{edfigure}

\begin{edfigure}[!tbp]
  \centering
  \includegraphics[width=\textwidth]{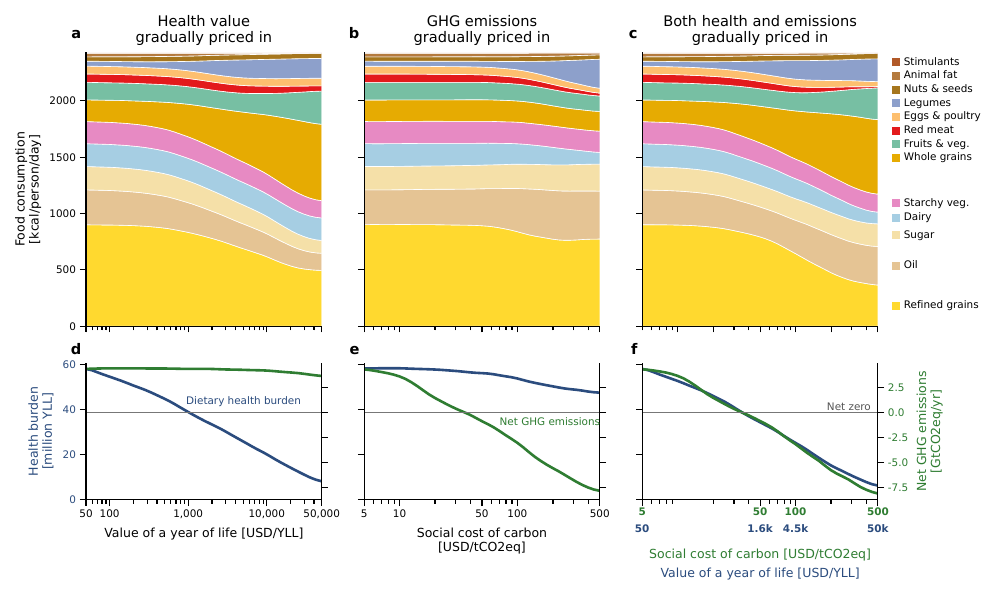}
  \caption{\textbf{Combined policy sensitivity under the calibrated deviation penalty.}
    Surrogate-median food consumption (top row), dietary-risk health burden, and net food system emissions (bottom row) as the value of a year of life, the social cost of carbon, or both policy levers vary.
    Non-policy uncertainty parameters are marginalized with Monte Carlo samples from the global-sensitivity-analysis design distribution.
    Because the global sensitivity analysis surrogate is trained over strictly positive policy ranges, single-policy sweeps hold the other policy lever at its lower global sensitivity analysis bound.
    Equivalent panels under loose and tight deviation-penalty calibrations are reported in the Supplementary Information.}
  \label{fig:eda-combined-sensitivity}
\end{edfigure}

\begin{edfigure}[thp]
  \centering
  \includegraphics[width=0.5\textwidth]{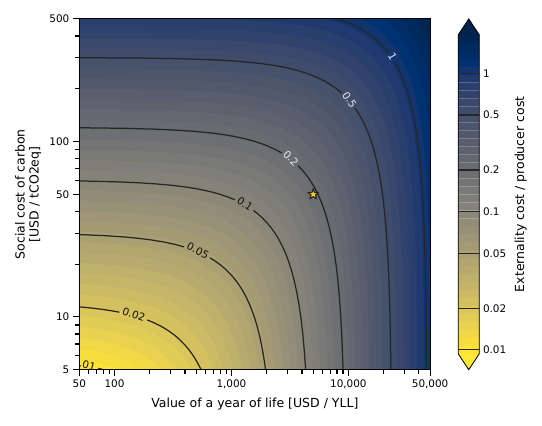}
  \caption{\textbf{Unpriced food-system externalities relative to today's food cost.}
    Single contour panel sharing the (value of a year of life, social cost of carbon) log-axes of \cref{fig:diet-health}, with agriculture and diets pinned to the GLADE reference scenario representing today's state.
    The color and contour values are the ratio of the externality cost --- the realized dietary-risk health burden and net food system emissions, valued at the operating point's own (YLL, social cost of carbon) valuations --- to the producer cost we pay for food today.
    The surface is rendered on a log scale (the ratio spans roughly two orders of magnitude across the displayed range) but contour and colorbar labels are the plain ratios.
    The gold star marks the central healthy-and-sustainable operating point used throughout the paper (YLL = \qty{\centralYllValue}{USD\per YLL}, carbon cost = \qty{\centralGhgPrice}{USD\per\tonne\COtwoeq}).}
  \label{fig:eda-today-externality-ratio}
\end{edfigure}

\begin{edfigure}[!tbp]
  \centering
  \includegraphics[width=\textwidth]{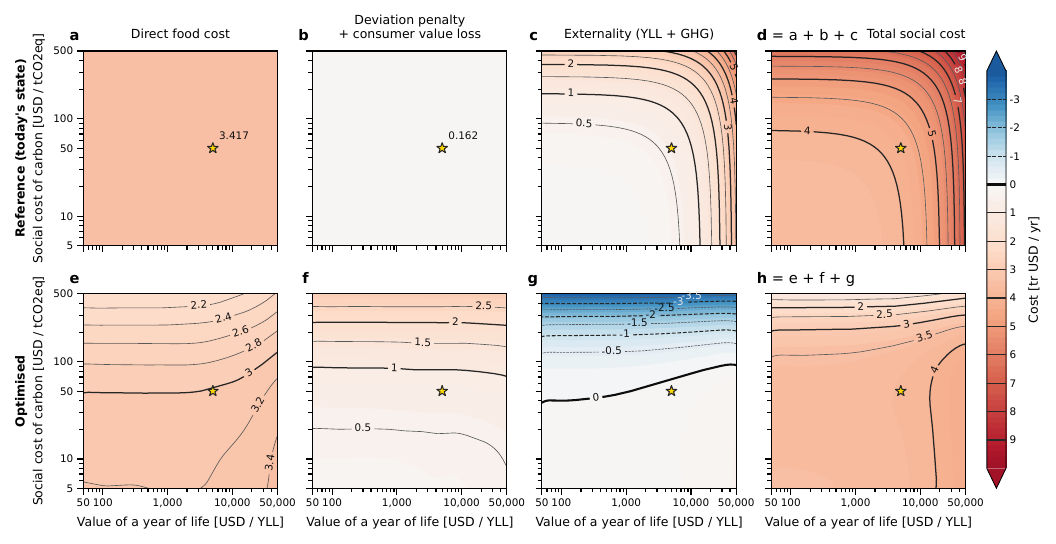}
  \caption{\textbf{Cost decomposition over the (value of a year of life, social cost of carbon) plane: reference (today's state) versus the optimized food system.}
    Eight contour panels in a 2$\times$4 grid sharing the axes of \cref{fig:diet-health}a,d.
    Top row: reference system (\texttt{scen-reference} -- diet and agriculture pinned to today's state, so the direct food cost and deviation column are constant across the plane and the consumer-value difference vanishes by construction).
    Bottom row: optimized system (GSA-surrogate medians; the cost components shift with the policy lever).
    Columns are, left to right: direct food cost (sum of surrogate-median producer subcategories from the model's objective breakdown: crop and animal production, fertilizer, trade, processing, consumption, feed conversion, land use, and exogenous feed supply); the production-stability deviation penalty (the L1 anchor on land use and production patterns) plus the change in calibrated consumer-value cost relative to the reference diet ($\mathrm{cv}_\textrm{opt} - \mathrm{cv}_\textrm{ref}$, with consumer-values entering the objective as negative utilities so that less-preferred optimized diets add a positive cost); externality cost evaluated at each operating point's own (YLL, social cost of carbon) valuations -- \(\textrm{YLL}\times \mathrm{YLL\,value} + (\textrm{CO}_2{+}\textrm{CH}_4{+}\textrm{N}_2\textrm{O})\times \mathrm{carbon\,cost}\); and total social cost, the cell-wise sum of the preceding three columns (\textbf{d}~=~\textbf{a}+\textbf{b}+\textbf{c}; \textbf{h}~=~\textbf{e}+\textbf{f}+\textbf{g}).
    The central healthy-and-sustainable operating point is marked with a star in each panel; all panels share a single diverging colorbar, with per-panel minor contour lines drawn where the panel's range warrants finer resolution than the major step.}
  \label{fig:eda-cost-surfaces}
\end{edfigure}

\begin{edfigure}[!tbp]
  \centering
  \includegraphics{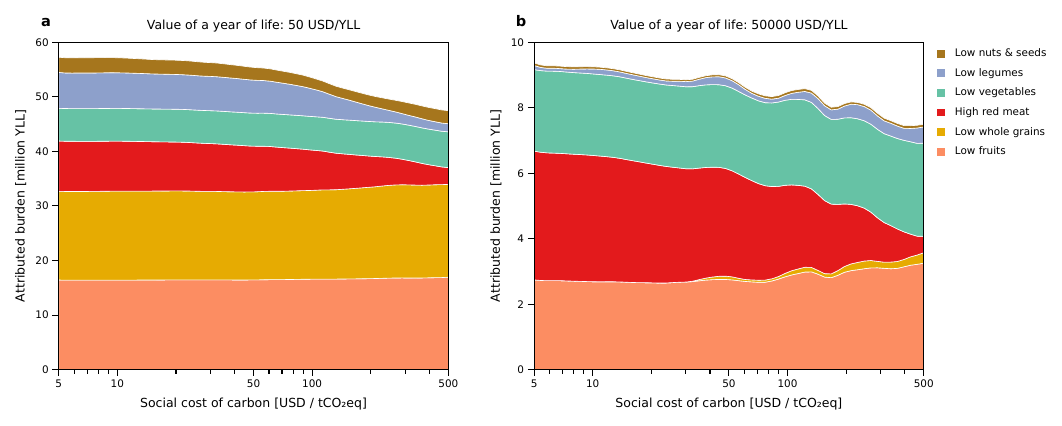}
  \caption{\textbf{Attribution of dietary-risk health burden at the two extremes of the value of a year of life.}
    Surrogate-median dietary-risk burden as a function of the social cost of carbon, stacked by attributable food-group risk factor, along \textbf{a} the low-YLL slice (\qty{50}{USD\per YLL}, matching panel e of \cref{fig:diet-health}) and \textbf{b} the high-YLL slice (\qty{50000}{USD\per YLL}, matching panel f).
    Attribution follows the same excess log-relative-risk allocation as \cref{fig:eda-burden-decomposition}.
    On the low-YLL slice \textbf{a}, the drop from \qty{\sim \yllSynergyLowGhg}{} to \qty{\sim \yllSynergyHighGhg}{MYLL\per yr} as the social cost of carbon rises from \qty{\lowGhgRefPrice}{} to \qty{\highGhgRefPrice}{USD\per\tonne\COtwoeq} is concentrated in two risk factors --- high red meat (largest absolute drop) and low legumes --- while low fruits, low vegetables and low whole grains, together accounting for a large share of the baseline burden, remain essentially flat (low whole grains, if anything, edges up slightly).
    This is the mechanism behind the region of synergy in \cref{fig:diet-health}a: a rising social cost of carbon displaces animal-source foods and oils primarily with legumes (\cref{fig:diet-health}e), and these three modeled risk factors move favorably in concert.
    The high-YLL slice \textbf{b} starts from an already much smaller residual burden because most of the avoidable burden has been eliminated by the value of a year of life alone; the remaining sensitivity to the social cost of carbon comes overwhelmingly from finishing the elimination of the red-meat tail.}
  \label{fig:eda-burden-attribution}
\end{edfigure}

\begin{edfigure}[!tbp]
  \centering
  \includegraphics[width=\textwidth]{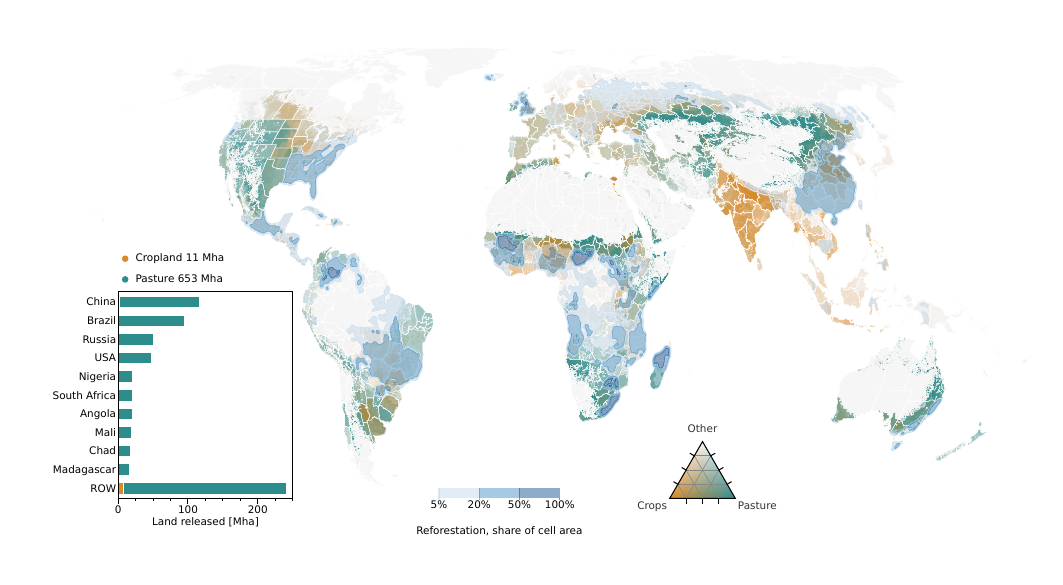}
  \caption{\textbf{Cropland and pasture utilization under the central healthy-and-sustainable scenario.}
    Bivariate cropland\,$\times$\,pasture utilization per cell under GLADE's central scenario (YLL = \qty{\centralYllValue}{USD\per YLL}, carbon cost = \qty{\centralGhgPrice}{USD\per\tonne\COtwoeq}), with the same ternary encoding as Figure~\ref{fig:overview-map}a.
    Blue contour bands mark the smoothed-density share of agricultural land released back to natural cover relative to the 2020 reference.
    The bar inset (lower left) ranks the top reforestation countries by released area, split into cropland (orange) and pasture (teal) source; global released-area totals (\qty{\croplandReleasedCentral}{Mha} cropland and \qty{\pastureReleasedCentral}{Mha} pasture) are annotated above the bars.}
  \label{fig:eda-central-scenario-map}
\end{edfigure}

\begin{edfigure}[!tbp]
  \centering
  \includegraphics[width=\textwidth]{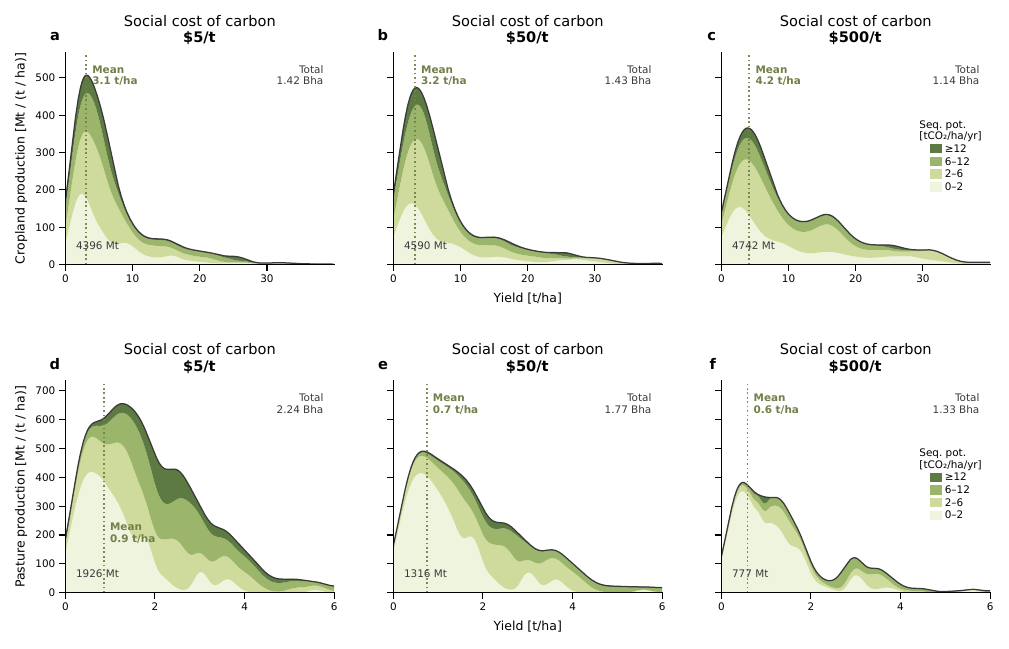}
  \caption{\textbf{Production-weighted distributions of cropland and pasture yields under a constant diet.}
    Detailed companion to the land-use row of \cref{fig:diet-abatement} (panels c, d), which collapses the same shift to area-by-yield-bucket bands.
    Production-weighted distributions of cropland \textbf{a}--\textbf{c} and pasture \textbf{d}--\textbf{f} yields at three representative carbon costs (\qty{\lowGhgRefPrice}{}, \qty{\centralGhgPrice}{} and \qty{\highGhgRefPrice}{USD\per\tonne\COtwoeq}, increasing left to right within each row) under a constant diet, showing increased utilization of high-yield cropland and a vast reduction in overall pasture area as the social cost of carbon rises.
    Stacked areas split each distribution by land carbon sequestration potential, showing that primarily pasture with a high sequestration potential is released and reforested.
    The panels also indicate total land use, total crop / grass production and overall area-weighted mean yields.}
  \label{fig:eda-yield-distributions}
\end{edfigure}

\begin{edfigure}[!tbp]
  \centering
  \includegraphics[width=\textwidth]{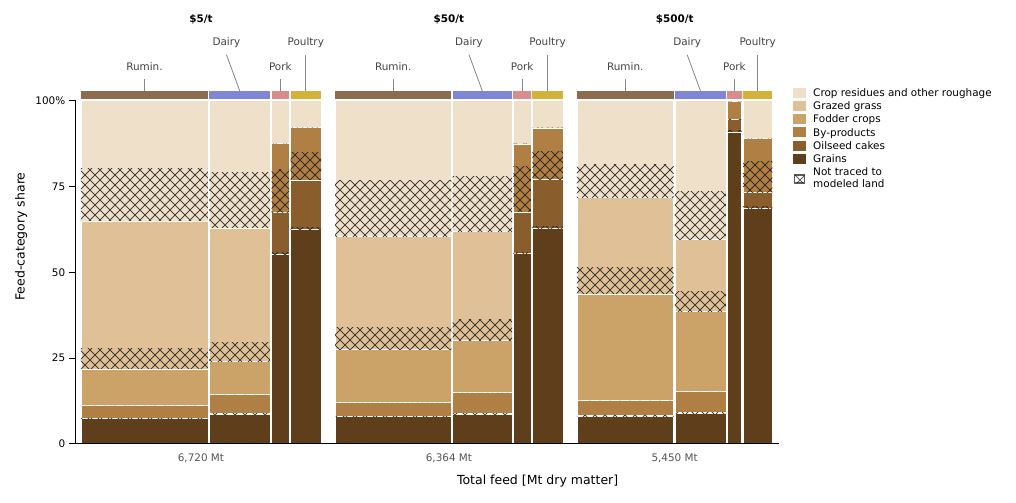}
  \caption{\textbf{Mosaic decomposition of global animal-feed intake under a constant diet.}
    Detailed companion to the feed-composition panel of \cref{fig:diet-abatement} (panel e), which collapses the same shift to a single continuous stacked area.
    Mosaic (Marimekko) decomposition of global animal-feed intake by source category at three representative carbon costs (\qty{\lowGhgRefPrice}{}, \qty{\centralGhgPrice}{} and \qty{\highGhgRefPrice}{USD\per\tonne\COtwoeq}): within each price block, columns are the four animal classes (ruminant meat, dairy, pork, poultry) with width set by their share of total feed intake, and the height split shows that class's feed composition.
    Feed categories are shaded and stacked by metabolisable-energy density, matching \cref{fig:diet-abatement} (panel e): the darkest browns mark the most energy-dense feeds at the base of each column.
    Cross-hatched sub-areas mark feed not traced to a modelled land or crop source (the landless calibration backstop).
    Total feed dry matter (\unit{\mega\tonne}) at each price is annotated below.}
  \label{fig:eda-feed-mosaic}
\end{edfigure}

\begin{edfigure}[!tbp]
  \centering
  \includegraphics[width=\textwidth]{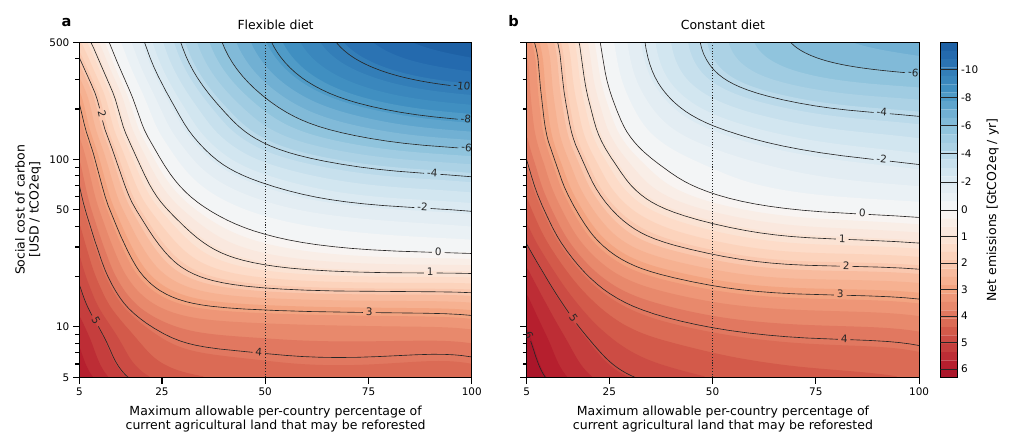}
  \caption{\textbf{Net food system emissions across the per-country reforestation cap.}
    Surrogate-median net food system emissions (\ch{CO2} + \ch{CH4} + \ch{N2O} in GWP\textsubscript{100}-equivalent \ch{CO2}; same diverging colormap and contour levels as \cref{fig:diet-health}d) as a function of the maximum fraction of each country's current agricultural land that may be reforested (x-axis, \qtyrange{5}{100}{\%}) and the social cost of carbon (y-axis), for \textbf{a},~flexible and \textbf{b},~constant diets.
    Medians are taken over the global-sensitivity-analysis ensemble with the value of a year of life held at zero, so each surface isolates the reforestation and carbon-pricing response.
    Loosening the cap deepens the achievable carbon sink, but for both diets a sufficiently high social cost of carbon still drives net food system emissions below zero even at a cap of 25\% --- the constant diet simply requires a higher carbon cost to do so.
    The paper's central configuration uses a \qty{50}{\%} cap.}
  \label{fig:eda-reforestation-surfaces}
\end{edfigure}

\begin{edtable}[!tbp]
  \centering
  \footnotesize
  \caption{\textbf{Uncertain parameters used in the global sensitivity analysis.}
    Ranges are either 90\% confidence intervals of a truncated normal distribution or the support of a uniform distribution, as indicated.
    The two GBD quantile parameters are active only in the flexible-diet ensemble; the remaining six are active in both ensembles.}
  \label{tab:eda-uncertain-parameters}
  \begin{tabular}{@{}llll@{}}
    \toprule
    Parameter & Distribution & Range / 90\% CI & Source of uncertainty \\
    \midrule
    Crop yield                             & Uniform          & 0.8--1.2 & GAEZ attainable-yield NRMSE~\cite{fao/iiasa-2023} \\
    \ch{CH4} emission factor               & Normal (90\% CI) & 0.5--1.5 & IPCC EF\,$\pm 40\%$ $\oplus$ GWP\textsubscript{100}\,$\pm 40\%$~\cite{ipcc-2019} \\
    \ch{N2O} emission factor               & Normal (90\% CI) & 0.3--1.7 & IPCC EF\,$\pm 50\%$ $\oplus$ GWP\textsubscript{100}\,$\pm 47\%$~\cite{ipcc-2019} \\
    \ch{CO2} emission and sequestration factor & Normal (90\% CI) & 0.3--1.7 & AR6 WGIII\,$\pm 70\%$~\cite{ipcc-2022} \\
    Food loss \& waste                     & Uniform          & 0.7--1.3 & FAO/UNEP SDG 12.3.1 inter-estimate disagreement~\cite{kummu-demoel-ea-2012,unitednationsenvironmentprogramme-2024} \\
    Feed conversion ratio                  & Uniform          & 0.8--1.2 & Post-calibration feed conversion ratio residual; precedent $\pm 20\%$~\cite{alexander-brown-ea-2016,springmann-clark-ea-2018} \\
    Protective risk factors                & Uniform (0,1)    & 0--1     & GBD CI interpolation~\cite{zheng-afshin-ea-2022} \\
    Harmful risk factors                   & Uniform (0,1)    & 0--1     & GBD CI interpolation~\cite{zheng-afshin-ea-2022} \\
    \bottomrule
  \end{tabular}
\end{edtable}

\end{refsection}


\clearpage
\setcounter{page}{1}
\section*{Supplementary Information}

\begin{suppfigure}[h]
  \centering
  \includegraphics{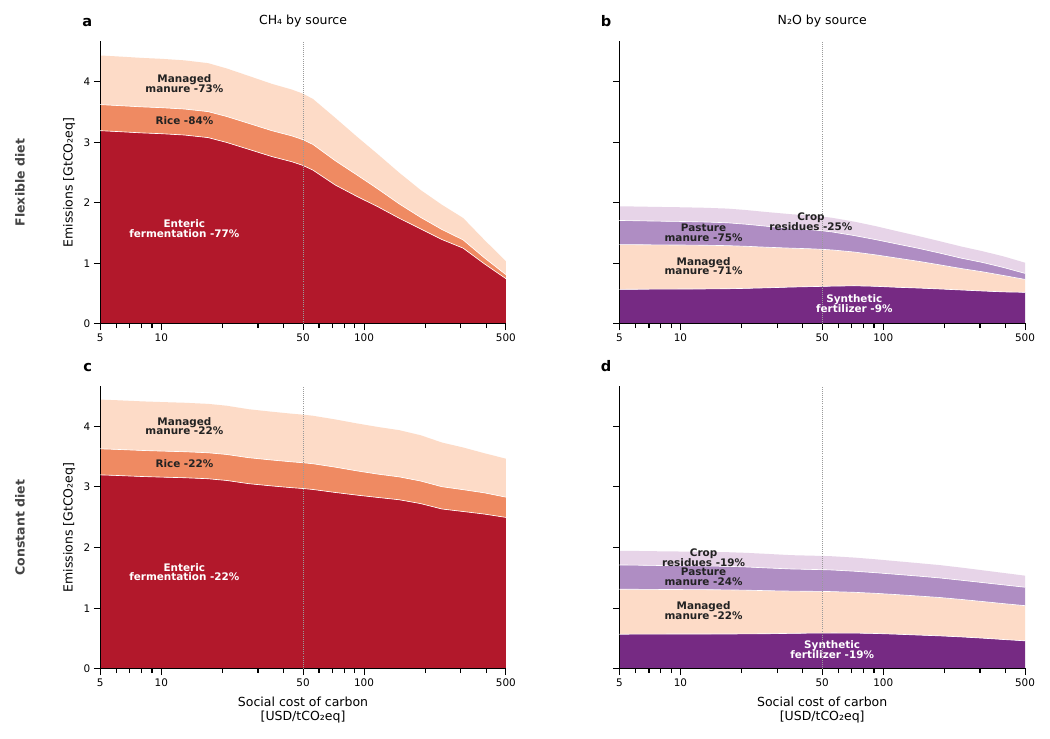}
  \caption[Non-\ch{CO2} emissions by source under flexible and constant diets]{\textbf{Methane and nitrous-oxide emissions by source under flexible and constant diets.}
    Companion to \cref{fig:diet-abatement}b: where that panel decomposes net emissions by gas at three reference carbon costs, here the two non-\ch{CO2} gases are decomposed by source category continuously over the social cost of carbon (x-axis, log scale; a faint guide marks the central reference carbon cost, with the low and high reference carbon costs at the axis endpoints).
    Gross \ch{CH4} (left column) and \ch{N2O} (right column) emissions are stacked by source, using the same source categories as the gross non-\ch{CO2} emissions bar of \cref{fig:overview-map}, with the flexible diet on the top row \textbf{a}, \textbf{b} and the constant diet on the bottom row \textbf{c}, \textbf{d}; the y-axis is shared down each gas column so the two diets are directly comparable, and each band is annotated with its net change across the sweep.
    Both diets are swept over the same social-cost-of-carbon grid with health left unvalued (value of a year of life set to zero), matching the flexible-diet definition of \cref{fig:diet-abatement}a,b, so the two rows differ only in whether the diet may re-optimize.
    Under a flexible diet the dominant sources collapse with rising carbon cost --- led by enteric fermentation and rice cultivation for \ch{CH4} and synthetic fertilizer for \ch{N2O} --- whereas under a constant diet the same sources abate only modestly through production-side reallocation.}
  \label{fig:si-emissions-by-source}
\end{suppfigure}

\begin{suppfigure}[p]
  \centering
  \includegraphics{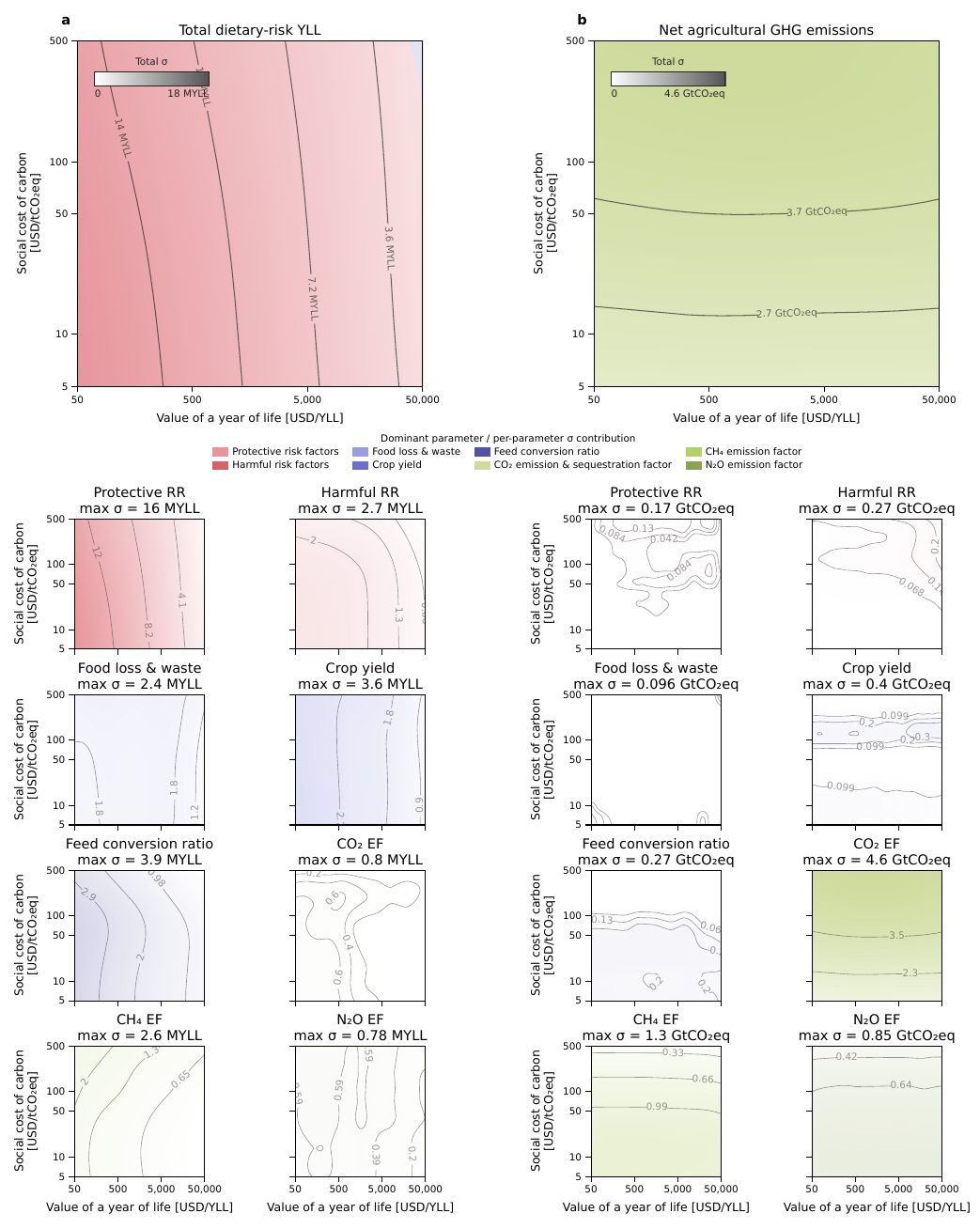}
  \caption[Drivers of outcome uncertainty across the full policy plane]{\textbf{Drivers of outcome uncertainty across the full policy plane.}
    Companion to Fig.~4: rather than slicing along a single policy axis, the conditional Sobol standard-deviation contributions are evaluated over the entire (value of a year of life, social cost of carbon) plane, with axes following Fig.~2 (x~=~value of a year of life, y~=~social cost of carbon, both log-scaled).
    The left column is total dietary-risk years of life lost; the right column is net food system emissions.
    \textbf{a}, \textbf{b},~Phase diagrams in which hue encodes the dominant non-policy parameter at each point of the plane and brightness encodes the magnitude of the total conditional standard deviation of the outcome (lighter~=~smaller, darker~=~larger; the gray strip gives the scale in the outcome's own units).
    The remaining panels show each non-policy parameter's conditional standard-deviation contribution as a separate surface over the same plane, on a shared per-column scale.
    All quantities are evaluated on the XGBoost surrogate and smoothed with a thin-plate-spline interpolant.
    Dietary-risk YLL uncertainty is governed by the protective dietary relative-risk factors and the feed-conversion ratio, while GHG-emissions uncertainty is dominated by the \ch{CO2} emission and sequestration factor across most of the high-social-cost-of-carbon region, consistent with the single-axis slices in Fig.~4.}
  \label{fig:si-phase-sobol}
\end{suppfigure}

\begin{suppfigure}[thp]
  \centering
  \includegraphics{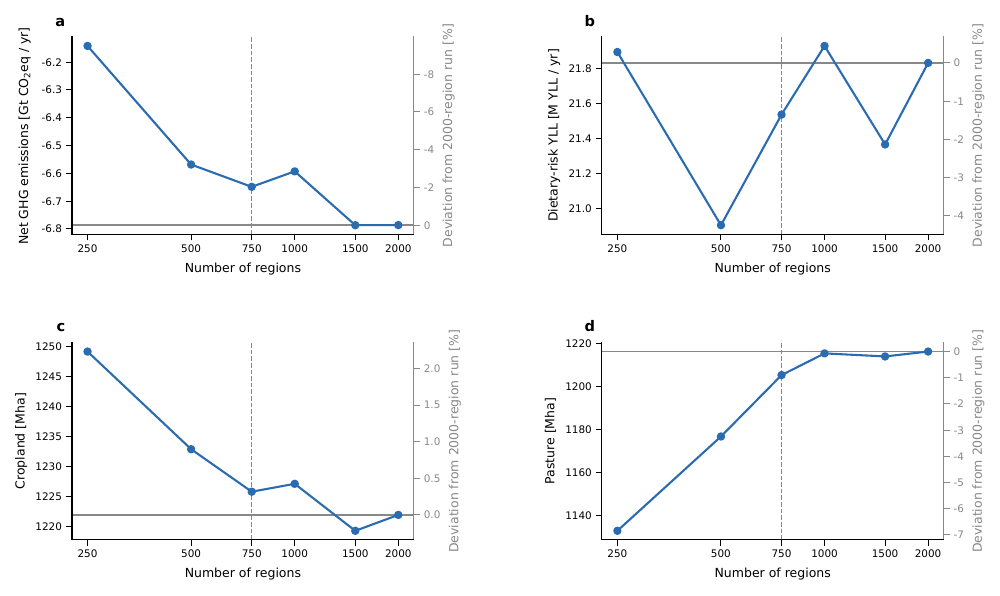}
  \caption[Sensitivity of the central operating point to spatial resolution]{\textbf{Sensitivity of the central healthy-and-sustainable point to spatial resolution.}
    The clustering step that aggregates first-level administrative units into model regions is rerun at six target counts (250, 500, 750, 1000, 1500, 2000); for each, the model is rebuilt and resolved at the central operating point (YLL = \$\centralYllValue, carbon cost = \$\centralGhgPrice / tCO\textsubscript{2}eq).  Panels show net food system emissions \textbf{a}, dietary-risk years of life lost \textbf{b}, cropland area \textbf{c}, and pasture area \textbf{d} on the left axis, with deviation from the highest-resolution (2000-region) run on the right axis.  The dashed gray line marks the 750-region default used elsewhere in the paper, and the solid gray line marks the 2000-region reference (0\,\% deviation).
    Headline outcomes are insensitive to spatial resolution across the full 250--2000-region range: relative to the 2000-region reference, net emissions, dietary-risk YLL and cropland all stay within \num{1.3}\,\% and pasture within \num{2.2}\,\%, with no systematic trend in resolution.  The 750-region default agrees with the highest-resolution run to within \num{0.6}\,\% on every outcome except pasture (\num{2.1}\,\%), confirming it as a safe working compromise.}
  \label{fig:si-region-resolution}
\end{suppfigure}

\begin{suppfigure}[thp]
  \centering
  \includegraphics{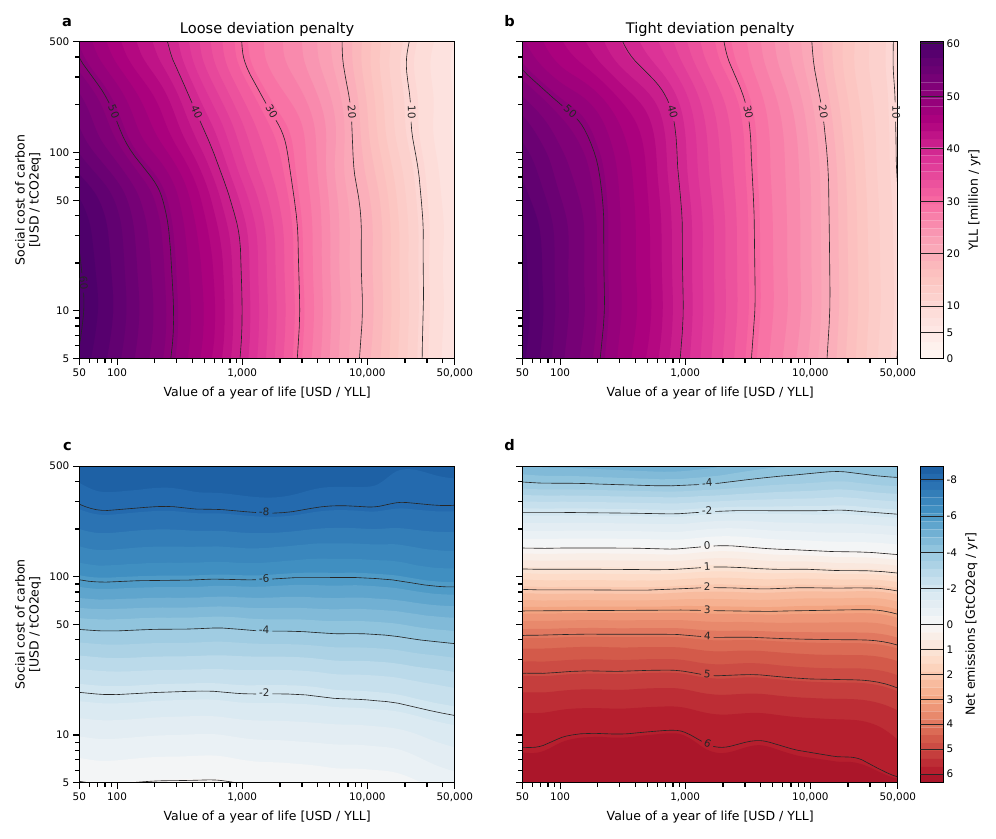}
  \caption[Sensitivity of the YLL and net-emissions surfaces to the deviation-penalty calibration]{\textbf{Sensitivity of the dietary-risk YLL and net-emissions surfaces to the deviation-penalty calibration.}
    Top row \textbf{a}, \textbf{b}: surrogate-median dietary-risk YLL across the (value of a year of life, social cost of carbon) grid; bottom row \textbf{c}, \textbf{d}: surrogate-median net food system emissions on the same grid.
    Left column shows the loose-penalty calibration; right column the tight-penalty calibration.
    Construction mirrors the calibrated surfaces shown in Fig.~2a (YLL) and Fig.~2d (net emissions): the median is taken over a Monte-Carlo ensemble of the six non-policy GSA parameters and Gaussian-smoothed on a dense log-grid.
    The qualitative result --- that the YLL surface is governed by the value of a year of life and the emissions surface by the social cost of carbon --- carries over to both calibrations; the chief quantitative effect of a tighter deviation penalty is to lift the entire emissions surface upward (compare \textbf{c} with \textbf{d}) so the net-zero locus sits at a higher social cost of carbon.}
  \label{fig:si-stability-contour-surfaces}
\end{suppfigure}

\begin{suppfigure}[thp]
  \centering
  \includegraphics{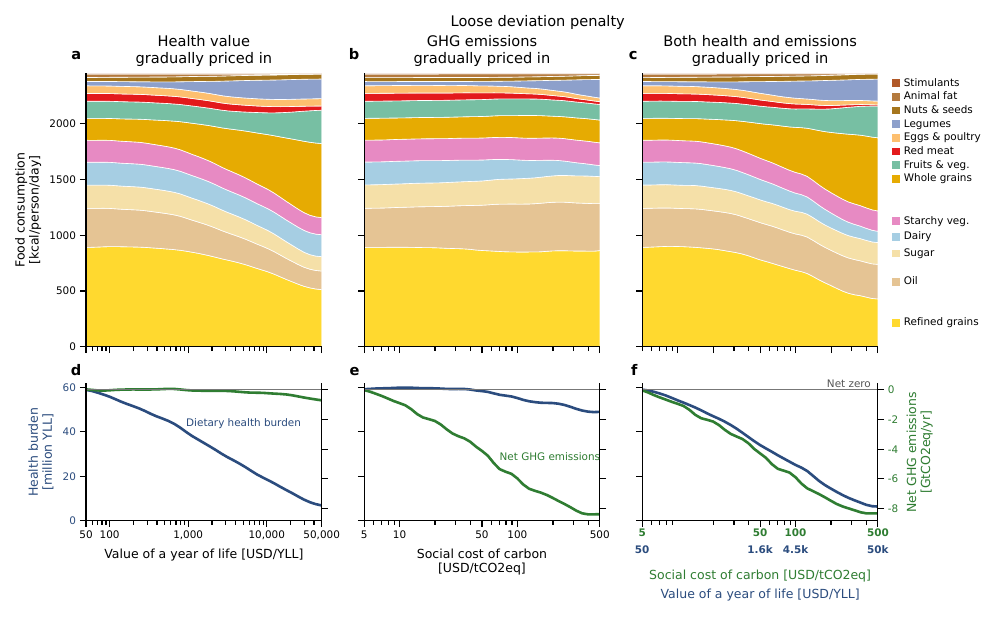}
  \caption[Combined policy sensitivity under a loose deviation penalty]{\textbf{Combined policy sensitivity under a loose deviation penalty.}
    Surrogate-median food consumption (top row), dietary-risk health burden, and net food system emissions (bottom row) as the value of a year of life, the social cost of carbon, or both policy levers vary, evaluated on the loose-penalty GSA surrogate (deviation-penalty coefficients reduced by $\sqrt{10}$ relative to the calibrated values).
    Non-policy uncertainty parameters are marginalized with Monte Carlo samples from the GSA design distribution; single-policy sweeps hold the other policy lever at its lower GSA bound.
    The calibrated case is shown in the Extended Data of the main paper.}
  \label{fig:si-combined-sensitivity-low}
\end{suppfigure}

\begin{suppfigure}[thp]
  \centering
  \includegraphics{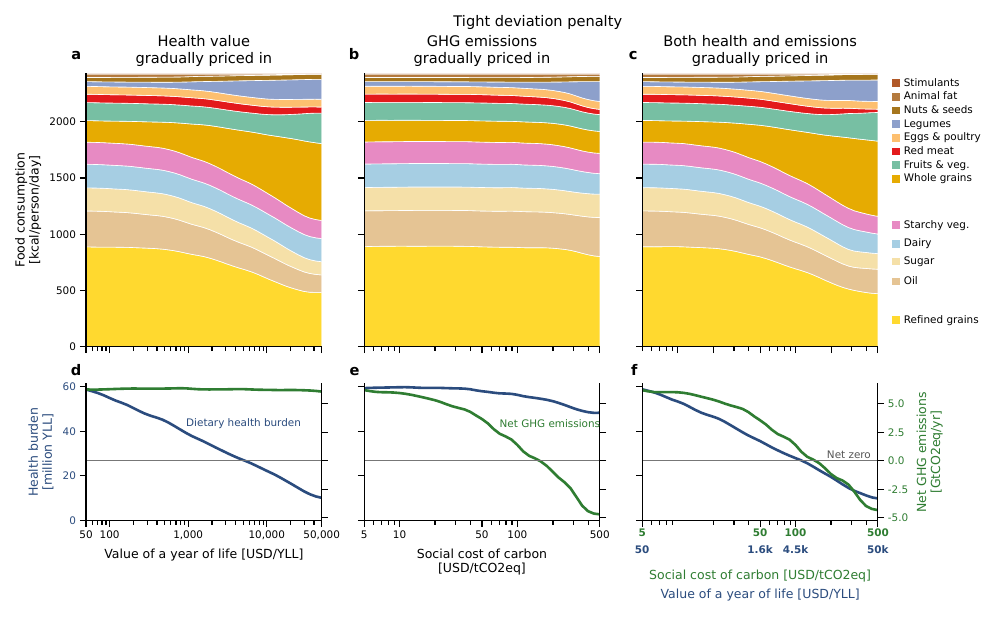}
  \caption[Combined policy sensitivity under a tight deviation penalty]{\textbf{Combined policy sensitivity under a tight deviation penalty.}
    Same as \cref{fig:si-combined-sensitivity-low}, but using the tight-penalty GSA surrogate (deviation-penalty coefficients increased by $\sqrt{10}$ relative to the calibrated values).}
  \label{fig:si-combined-sensitivity-high}
\end{suppfigure}

\end{document}